\documentclass[11pt]{article}
\usepackage{amsmath,amsthm,amssymb}
\usepackage{graphicx}
\usepackage{algorithm}
\usepackage[font={small,it}]{caption}
\usepackage{authblk}
\usepackage{subfig}

\author[1]{George Haller\thanks{Email address for correspondence: georgehaller@ethz.ch}}
\author[1]{Alireza Hadjighasem\thanks{alirezah@ethz.ch}}
\author[2]{Mohammad Farazmand\thanks{mohammad.farazmand@physics.gatech.edu}}
\author[3]{Florian Huhn\thanks{
florian.huhn@imes.mavt.ethz.ch}}

\affil[1]{\footnotesize Department of Mechanical and Process Engineering, ETH Z\"{u}rich, Leonhardstrasse 21, 8092 Zurich, Switzerland}

\affil[2]{\footnotesize Department of Mechanical Engineering, Massachusetts Institute of Technology, 77 Massachusetts Av., Cambridge, MA 02139-4307, USA}

\affil[3]{\footnotesize Institute of Aerodynamics and Flow Technology, German Aerospace Center, Bunsenstrasse 10, 37073 Gottingen, Germany}

\title{Defining Coherent Vortices Objectively from the Vorticity}
\date{}

\providecommand{\definitionname}{Definition}
\providecommand{\theoremname}{Theorem}

\newtheorem{thm}{\protect\theoremname}
\newtheorem{defn}{\protect\definitionname}

\begin{document}

\maketitle

\begin{abstract}
Rotationally coherent Lagrangian vortices are formed by tubes of deforming
fluid elements that complete equal bulk material rotation relative
to the mean rotation of the deforming fluid volume. We show that initial
positions of such tubes coincide with tubular level surfaces of the
Lagrangian-Averaged Vorticity Deviation (LAVD), the trajectory integral
of the normed difference of the vorticity from its spatial mean. LAVD-based
vortices are objective, i.e., remain unchanged under time-dependent
rotations and translations of the coordinate frame. In the limit of
vanishing Rossby numbers in geostrophic flows, cyclonic LAVD vortex
centers are precisely the observed attractors for light particles.
A similar result holds for heavy particles in anticyclonic LAVD vortices.
We also establish a relationship between rotationally coherent Lagrangian
vortices and their instantaneous Eulerian counterparts. The latter
are formed by tubular surfaces of equal material rotation rate, objectively
measured by the Instantaneous Vorticity Deviation (IVD). We illustrate
the use of the LAVD and the IVD to detect rotationally coherent Lagrangian
and Eulerian vortices objectively in several two- and three-dimensional
flows. 
\end{abstract}

\section{Introduction}

Coherent vortices still have no universal definition in fluid mechanics,
but two main features of a possible definition have been emerging.
First, vortices are broadly agreed to be concentrated regions of high
vorticity. Some authors require this dominance of vorticity relative
to other flow domains (McWilliams 1984, Hussein 1986), others expect
it relative to the strain in the same domain (Okubo 1970, Hunt et
al. 1988, Weiss 1991, Hua \& Klein 1998, Hua, McWilliams \& Klein
1998). Yet others compare vorticity to strain in the rate-of-strain
eigenbasis (Tabor \& Klapper 1995; see also Lapeyre, Klein \& Hua
1999, and Lapeyre, Hua \& Legras 2001).

Second, vortices are generally viewed as evolving domains with a high
degree of material invariance. Lugt (1979) writes that a vortex is
a ``multitude of material particles rotating around a common center''.
McWilliams (1984) expects the vortex to ``persist under passive advection
by the large-scale flow''. Chong, Perry \& Cantwell (1990) propose
to capture vortices by finding spiraling particle motions in the frozen-time
limit of the flow. Vortices are described as ``highly impermeable
to inward and outward particle fluxes'' by Provenzale (1999), who
requires small relative dispersion within vortex cores (see also Cucitore
Quadrio \& Baron 1999). Chakraborty, Balachandar \& Adrian (2005)
proposes that both swirling motion and small particle separation should
be distinguishing features of a vortex core. Haller (2005) views vortices
as sets of trajectories with a persistent lack of Lagrangian hyperbolicity.
Chelton et al. (2011) observe that nonlinear eddies (vortices with
a rotation speed exceeding their translation speed) trap fluid in
their interior and transport them along. Finally, in a similar geophysical
setting, Mason, Pascual \& McWilliams (2014) stress that vortices
are ``efficient carriers of mass and its physical, chemical, and
biological properties''.

The core of a coherent vortex is, therefore, broadly expected to be
an impermeable material region marked by a high concentration of vorticity.
What constitutes high vorticity is, however, subject to individual
judgement, thresholding, and choice of the reference frame. It is
therefore the material invariance of a vortex core that holds more
promise as a first requirement in an unambiguous vortex definition.
Indeed, the Lagrangian nature of a vortex can simply be assured by
defining its boundary as a tubular (i.e., cylindrical, cup-shaped
or toroidal) material surface. The challenging next step is then to
select such a material surface in a way that it also encloses a region
of concentrated vorticity.

Unlike vorticity, materially defined vortex boundary surfaces are
inherently frame-invariant, defined by a set of fluid trajectories
rather than by coordinates or instantaneous scalar field values. In
continuum mechanics terminology, a material vortex boundary must therefore
be objective, i.e., invariant with respect to all Euclidean frame
changes of the form 
\begin{equation}
x=Q(t)y+b(t),\label{eq:observer_change}
\end{equation}
where $x\in\mathbb{R}^{3}$ and $y\in\mathbb{R}^{3}$ denote coordinates
in the original and in the transformed frame, respectively; $Q(t)\in SO(3)$
is an arbitrary rotation matrix; and $b(t)\in\mathbb{R}^{3}$ is an
arbitrary translation vector (Truesdell \& Noll 1965). Paradoxically,
with the exception of the approach initiated by Tabor \& Klapper (1995),
none of the instantaneous Eulerian vortex criteria listed above are
objective. Accordingly, they may only detect coherent structures after
passage to an appropriately rotating or translating coordinate frame.
For instance, the unsteady Navier-Stokes velocity field 
\begin{equation}
v(x,t)=\left(\begin{array}{c}
x_{1}\sin4t+x_{2}(2+\cos4t)\\
x_{1}(\cos4t-2)-x_{2}\sin4t\\
0
\end{array}\right),
\label{eq:rotating_saddle_flow}
\end{equation}

is classified as a vortex by the Okubo--Weiss, Hua--Klein, Hua--McWilliams--Klein,
and Chakraborty--Balachandar--Adrian criteria, as well as by the $Q$-criterion
of Hunt et al., the $\Delta$-criterion of Chong, Perry and Cantwell,
and the nonlinear eddy criterion of Chelton et al. In reality, \eqref{eq:rotating_saddle_flow}
is a rotating saddle-point flow, with typical trajectories growing
exponentially in norm. This instability, however, only becomes detectable
to these criteria after one passes to an appropriately chosen rotating
frame (Haller 2005, 2015). Promisingly, if we simply impose the localized
high-vorticity requirement, the constant-vorticity flow \eqref{eq:rotating_saddle_flow}
is immediately discounted as a vortex without further need for analysis.

Selecting vortex boundaries as material surfaces ensures material
invariance for the vortex, but any tubular material surface can a
priori be considered for this purpose. Recent stretching-based variational
approaches narrow down this consideration to exceptional material
tubes that remain perfectly unfilamented under material advection
(Haller \& Beron--Vera 2013, Blazevski \& Haller 2004, Haller 2015).
As an alternative, Farazmand \& Haller (2016) seek vortex boundaries
as maximal material tubes along which material elements complete the
same polar rotation over a finite time interval of interest. These
approaches have proven effective in two-dimensional flows. They, however,
rely on a precise computation of the invariants of the Cauchy--Green
strain tensor along a Lagrangian grid, which requires the accurate
numerical differentiation of trajectories with respect to their initial
positions. This presents a challenge in three-dimensional unsteady
flows, for which the polar rotation approach additionally fails to
be objective. Most importantly, however, Lagrangian strain-based approaches
offer no link between material vortices and the expected high vorticity
concentration, a defining feature of observed vortices.

In summary, despite recent advances in vortex criteria and fluid trajectory
stability analysis, a fully three-dimensional, computationally tractable
and objective global vortex definition, with guaranteed material invariance
and experimentally observable rotational coherence, has not yet emerged.
Here we propose such a vortex definition and a corresponding vortex
detection technique.

Our approach is based on a recently obtained, unique decomposition
of the deformation gradient into the product the two deformation gradients:
one for a purely straining flow and one for a purely rotational flow
(Haller 2016). This rotational deformation gradient, the dynamic rotation
tensor, obeys the temporal superposition property of rigid body rotations,
thereby eliminating a dynamical inconsistency of the classic polar
rotation tensor used in classical continuum mechanics. The dynamic
rotation tensor can further be factorized into a spatial mean-rotation
component and a deviation from this rotation. The latter deviatory
part yields an objective, intrinsic material rotational angle relative
to the deforming fluid mass.

We then define a rotationally coherent Lagrangian vortex as a nested
set of material tubes, each exhibiting uniform intrinsic material
rotation. Such a vortex turns out to be foliated by outward decreasing
tubular level sets of the Lagrangian-averaged vorticity (LAVD). Additionally,
we prove that the center of an LAVD-based vortex is always the observed
attractor for nearby finite-size (inertial) particle motions in geostrophic
flows.

In the limit of zero advection time, our Lagrangian vortex definition
turns into an objective Eulerian vortex definition: a set of tubular
surfaces of equal intrinsic rotation rate. These surfaces are tubular
level sets of the instantaneous vorticity deviation (IVD), providing
a mathematical link between rotationally coherent Eulerian and Lagrangian
vortices: the former are effectively derivatives of the latter. We
illustrate our results on several examples, ranging from analytic
velocity fields to time-dependent two- and three-dimensional models
and observational data.

\section{Set-up}

We consider an unsteady velocity field $v(x,t)$, defined on a possibly
time-dependent spatial domain $U(t)\subset\mathbf{\mathbb{R}}^{3}$
over a finite time interval $[t_{0},t_{1}].$ We assume that $U(t)$
is invariant under the fluid flow generated by the velocity field
(cf. eq. \eqref{eq:invariance_of_U} below). Thus, $U(t)$ is either
a physical domain with an impermeable boundary, or $U(t)$ is a material
domain formed by a set of evolving trajectories of $v(x,t)$.

We write the velocity gradient $\nabla v$ as 
\begin{equation}
\nabla v(x,t)=D(x,t)+W(x,t),\label{eq:gradvdecomp}
\end{equation}
where $D=\frac{1}{2}\left(\nabla v+\left[\nabla v\right]^{T}\right)$
is the rate of stain tensor and $W=\frac{1}{2}\left(\nabla v-\left[\nabla v\right]^{T}\right)$
is the spin tensor. We recall that the vorticity $\omega=\nabla\times v$
of the fluid satisfies 
\begin{equation}
We=-\frac{1}{2}\omega\times e,\qquad\forall e\in\mathbb{R}^{3}.\label{eq:vortdef}
\end{equation}
We will also use the instantaneous spatial mean $\bar{\omega}$ of
the vorticity over $U(t)$, defined as 
\begin{equation}
\bar{\omega}(t)=\frac{\int_{U(t)}\omega(x,t)\,dV}{\mathrm{vol}\,(U(t))},\label{eq:mean_vorticity}
\end{equation}
where $\mathrm{vol}\,(\cdot\,)$ denotes the volume for three-dimensional
flows, and the area for two-dimensional flows. Accordingly, $dV$
refers to the volume or area element, respectively, in $U(t)$.

Under general observer changes of the form \eqref{eq:observer_change},
the spin tensor and the vorticity in the new $y$ coordinate frame
take the form 
\begin{equation}
\tilde{W}(y,t)=Q^{T}(t)W(x,t)Q(t)-Q^{T}(t)\dot{Q}(t),\qquad\tilde{\omega}(y,t)=Q^{T}(t)\omega(x,t)+\dot{q}(t),\label{eq:frame-dependence}
\end{equation}
with the vector $\dot{q}$ defined uniquely by the relation $\dot{Q}Q^{T}e=\frac{1}{2}\dot{q}\times e$
for all $e\in\mathbb{R}^{3}$ (see, e.g., Truesdell \& Rajagopal 2009).
Formula \eqref{eq:frame-dependence} shows that the spin tensor and
the vorticity are not objective quantities: the eigenvalues and eigenvectors
of $W$ change in rotating frames, and so does the direction and the
magnitude of $\omega$. Thus, neither $W$ nor $\omega$ is, by itself,
suitable for defining distinguished \emph{material} sets in the flow.
This is because material sets are tied to evolving fluid particles
without any reference to coordinates, and hence are inherently frame-invariant.
More generally statements about the material response of a moving
continuum cannot depend on the observer and hence should be objective
(Gurtin 1982).

Fluid particle trajectories generated by $v(x,t)$ are solutions of
the differential equation 
\[
\dot{x}=v(x,t),
\]
defining the flow map

\begin{equation}
\mathcal{F}_{t_{0}}^{t}\colon x_{0}\mapsto x(t;x_{0}),\qquad\ t\in[t_{0},t_{1}],\label{eq:flow map}
\end{equation}
as the mapping from initial particle positions $x_{0}\in U(t_{0})$
to their later positions $x(t;x_{0})\in U(t)$. The assumed invariance
of $U(t)$ over the time interval $[t_{0},t_{1}]$ can now be conveniently
expressed as 
\begin{equation}
\mathcal{F}_{t_{0}}^{t}(U(t_{0}))=U(t)\qquad t\in[t_{0},t_{1}].\label{eq:invariance_of_U}
\end{equation}

We will also use the notion of a \emph{material surface}, which is
a smooth, codimension-one, time-dependent surface family $\mathcal{M}(t)$
advected by the flow, i.e., $\mathcal{M}(t)=\mathcal{F}_{t_{0}}^{t}(\mathcal{M}(t_{0}))\qquad t\in[t_{0},t_{1}].$

The deformation gradient 
\begin{equation}
F_{t_{0}}^{t}(x_{0})=\nabla\mathcal{F}_{t_{0}}^{t}(x_{0})\label{eq:defgradient}
\end{equation}
is a linear map, taking initial infinitesimal perturbations to the
fluid trajectory $x(t;x_{0})$ at time $t_{0}$ to their later positions
at time $t$. Although often believed otherwise, $F_{t_{0}}^{t}$
is \emph{not} objective: its eigenvectors and eigenvalues depend on
the frame of reference (see, e.g., Liu 2004). Therefore, the invariants
of $F_{t_{0}}^{t}$ do not provide an objective indication of the
rotational component of the deformation.

\section{Finite material rotation from the Dynamic Polar Decomposition}

We seek to identify coherent Lagrangian vortices as the union of tubular
material surfaces in which fluid elements exhibit the same bulk material
rotation over a finite time interval of interest. Individual material
fibers based at an initial point $x_{0}$ in a deforming continuum,
however, all rotate around different axes and by different angles.
In recent work (Farazmand \& Haller 2016), we used the classic polar
rotation angle (PRA) from finite strain theory to identify pointwise
bulk material rotation in a moving fluid systematically. The use of
the PRA, however, also leaves several challenges unaddressed, as we
discuss in Appendix A. Most notable of these are an inconsistency
of the PRA with experimentally observed dynamic rotation angles of
spherical tracers in fluids, and its lack of objectivity in three
dimensions.

To address these challenges, we use here the recently developed \emph{Dynamic
Polar Decomposition} (DPD) to identify a dynamically consistent and
fully frame-invariant rotational component in the finite deformation
of fluid elements (Haller 2016). This decomposition gives a unique,
time-evolving factorization of the deformation gradient into the product
of two deformation gradients: one for a purely rotational flow with
zero rate of strain, and one for a purely straining flow with zero
vorticity.

Specifically, the unique right DPD of $F_{t_{0}}^{t}$ at $x_{0}$
can be written as 
\begin{equation}
F_{t_{0}}^{t}=O_{t_{0}}^{t}M_{t_{0}}^{t},\qquad t\in[t_{0},t_{1}],\label{eq:DPD-1}
\end{equation}
where the proper orthogonal \emph{dynamic rotation tensor} $O_{t_{0}}^{t}=\partial_{a_{0}}a(t)$
is the deformation gradient of the purely rotational flow 
\begin{equation}
\dot{a}=W\left(x(t;x_{0}),t\right)a,\label{eq:Odef}
\end{equation}
and the non-degenerate \emph{right dynamic stretch tensor} $M_{t_{0}}^{t}=\partial_{b_{0}}b(t)$
is the deformation gradient of the purely straining flow 
\begin{equation}
\dot{b}=O_{t}^{t_{0}}D\left(x(t;x_{0}),t\right)O_{t_{0}}^{t}b.\label{eq:Mdef}
\end{equation}
The linear velocity field for $a(t)$ in \eqref{eq:Odef} is strainless
because its coefficient matrix $W$ is skew-symmetric. Similarly,
the linear velocity field for $b(t)$ in \eqref{eq:Mdef} is irrotational
because its coefficient matrix $O_{t}^{t_{0}}D\left(x(t;x_{0}),t\right)O_{t_{0}}^{t}$
is symmetric. Unlike the classic polar rotation tensor (cf. Appendix
A), the dynamic rotation tensor $O_{t_{0}}^{t}$ is \emph{dynamically
consistent}, i.e., satisfies the fundamental superposition property
of solid-body rotations: 
\begin{equation}
O_{t_{0}}^{t}=O_{s}^{t}O_{t_{0}}^{s},\qquad s,t\in[t_{0},t_{1}].\label{eq:groups}
\end{equation}
This follows because $O_{t_{0}}^{t}$ is the fundamental matrix solution
of a classical linear differential equation and hence satisfies the
process property noted in \eqref{eq:groups} (cf. Dafermos 1971, Arnold
1978). In contrast, $M_{t_{0}}^{t}$ is the fundamental matrix solution
of a non-classical linear differential equation with memory, i.e.,
with explicit dependence on the initial time $t_{0}$. Such fundamental
solutions do not obey the process-property indicated in \eqref{eq:groups}.
The reason behind the dynamical inconsistency \eqref{eq:no_product}
of polar rotations is a similar memory effect in eq. \eqref{eq:polar_rot_ODE}.

The decomposition in \eqref{eq:DPD-1} is a right-type decomposition,
i.e., the dynamic stretch tensor precedes the dynamic rotation tensor
from the right. Just as for the classic polar decomposition, a left-type
version of the DPD is also available (Haller 2016).

\section{Lagrangian-averaged vorticity deviation (LAVD)}

Despite its dynamical consistency, the dynamic rotation tensor $O_{t_{0}}^{t}$
is not objective. Its frame-dependence is the consequence of the frame-dependence
of the spin tensor $W(x,t)$ appearing in the differential equation
\eqref{eq:Odef}. The single remaining challenge out of those listed
in Appendix A is, therefore, to identify an objective part of the
rotation described by $O_{t_{0}}^{t}$ which also preserves the dynamical
consistency of $O_{t_{0}}^{t}$. Below we recall further results from
Haller (2016), and use them to address this challenge.

The dynamic rotation tensor $O_{t_{0}}^{t}$ can further be factorized
into two deformation gradients: one for a spatially uniformly rotating
flow, and one for a flow that describes deviations from this uniform
rotation. Specifically, we have 
\begin{equation}
O_{t_{0}}^{t}=\Phi_{t_{0}}^{t}\Theta_{t_{0}}^{t},\label{eq:decomp}
\end{equation}
where the proper orthogonal \emph{relative rotation tensor} $\Phi_{t_{0}}^{t}=\partial_{\alpha_{0}}\alpha(t)$
is dynamically consistent, serving as the deformation gradient of
the relative rotation flow 
\begin{equation}
\dot{\alpha}=\left[W\left(x(t;x_{0}),t\right)-\bar{W}\left(t\right)\right]\alpha.\label{eq:Odef-1}
\end{equation}
In contrast, the proper orthogonal \emph{mean rotation tensor} $\Theta_{t_{0}}^{t}=D_{\beta_{0}}\beta(t)$
is the deformation gradient of the mean-rotation flow 
\begin{equation}
\dot{\beta}=\Phi_{t}^{t_{0}}\bar{W}\left(t\right)\Phi_{t_{0}}^{t}\beta.\label{eq:Mdef-1}
\end{equation}
The mean rotation tensor $\Theta_{t_{0}}^{t}$ is not dynamically
consistent because \eqref{eq:Mdef-1} exhibits the same memory effect
discussed for \eqref{eq:Mdef}.

\begin{figure}
\centering 
\includegraphics[width=0.8\textwidth]{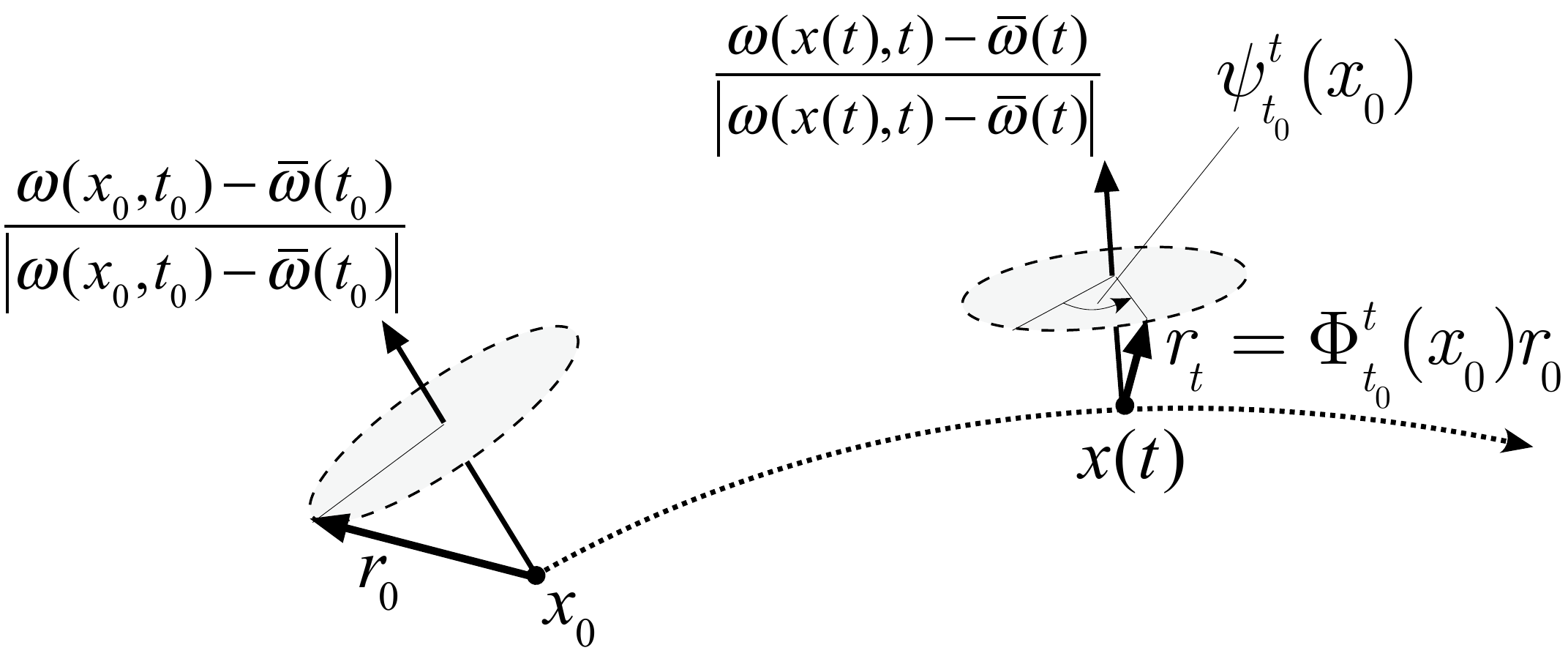}
\caption{The geometry of the intrinsic rotation angle $\psi_{t_{0}}^{t}(x_{0})$
along a material trajectory $x(t)$ in a three-dimensional deforming
continuum.}
\label{fig:intrinsic} 
\end{figure}

The dynamic consistency of $\Phi_{t_{0}}^{t}$ implies that the total
angle swept by this tensor around its own axis of rotation is dynamically
consistent. This angle $\psi_{t_{0}}^{t}(x_{0})$, called\emph{ intrinsic
rotation angle} (see Fig. \eqref{fig:intrinsic}), therefore satisfies
\[
\psi_{t_{0}}^{t}(x_{0})=\psi_{s}^{t}(x_{0})+\psi_{t_{0}}^{s}(x_{0}),\qquad s,t\in[t_{0},t_{1}].
\]

In addition, as shown in Haller (2016), $\psi_{t_{0}}^{t}(x_{0})$
is objective both in two and three dimensions. In two dimensions,
even the tensor $\Phi_{t_{0}}^{t}$ itself turns out to be objective,
not just its associated scalar field $\psi_{t_{0}}^{t}(x_{0})$.

Using the results obtained in Haller (2016)\emph{, }the intrinsic
dynamic rotation\emph{ $\psi_{t_{0}}^{t}(x_{0})$} can be computed
as 
\begin{equation}
\psi_{t_{0}}^{t}(x_{0})=\frac{1}{2}\mathrm{LAVD}_{t_{0}}^{t}(x_{0}),\label{eq:psidef}
\end{equation}
with the \emph{Lagrangian-Averaged Vorticity Deviation} (LAVD) defined
here as 
\begin{equation}
\mathrm{LAVD}_{t_{0}}^{t}(x_{0}):=\int_{t_{0}}^{t}\left|\omega(x(s;x_{0}),s)-\bar{\omega}(s)\right|\,ds.\label{eq:LAVD}
\end{equation}
The objectivity of $\psi_{t_{0}}^{t}$ and LAVD can be confirmed directly
from formula \eqref{eq:frame-dependence}. Indeed, under a Euclidean
observer change $x=Q(t)y+b(t)$, the transformed vorticity $\tilde{\omega}(y,t)$
satisfies 
\begin{eqnarray}
\left|\tilde{\omega}(y(s),s)-\tilde{\bar{\omega}}(s)\right| & = & \left|Q^{T}(s)\omega(x(s),s)+Q^{T}(t)\dot{q}(t)-\left(Q^{T}(s)\bar{\omega}(s)+Q^{T}(t)\dot{q}(t)\right)\right|\nonumber \\
 & = & \left|Q^{T}(s)\left[\omega(x(s),s)-\bar{\omega}(s)\right]\right|\nonumber \\
 & = & \left|\omega(x(s),s)-\bar{\omega}(s)\right|,\label{eq:objective}
\end{eqnarray}
because the rotation matrix $Q^{T}(s)$ preserves the length of vectors.
We summarize the results of this section in a theorem. \begin{thm}
For an infinitesimal fluid volume starting from $x_{0}$, the $\mathrm{LAVD}_{t_{0}}^{t}(x_{0})$
field is a dynamically consistent and objective measure of bulk material
rotation relative to the spatial mean-rotation of the fluid volume
$U(t)$. Specifically, $\mathrm{LAVD}_{t_{0}}^{t}(x_{0})$ is twice
the intrinsic dynamic rotation angle generated by the relative rotation
tensor $\Phi_{t_{0}}^{t}$. The latter tensor is obtained from the
dynamically consistent decomposition 
\begin{equation}
F_{t_{0}}^{t}=\Phi_{t_{0}}^{t}\Theta_{t_{0}}^{t}M_{t_{0}}^{t}\label{eq:full_decomp}
\end{equation}
with the deformation gradient $\Theta_{t_{0}}^{t}$ of a pure rigid-body
rotation, and with the deformation gradient $M_{t_{0}}^{t}$of a unique,
purely straining flow. \end{thm}

For detailed proofs of all statements in Theorem 1, we refer the reader
to Haller (2016). Importantly, this theorem enables the extraction
of an objective and dynamically consistent, material rotation component
from the deformation gradient without carrying out the differentiation
with respect to initial conditions in the definition \eqref{eq:defgradient}
of $F_{t_{0}}^{t}$.

\section{Rotationally coherent Lagrangian vortices}

We now use the LAVD to identify objectively material tubes along which
small fluid volumes experience the same bulk rotation over $[t_{0},t_{1}]$
relative to the mean rigid-body rotation of the fluid. By Theorem
1, the time $t_{0}$ positions of such material tubes are tubular
level surfaces of the scalar function $\mathrm{LAVD}_{t_{0}}^{t_{1}}(x_{0})$.
By a \emph{tubular set}, we mean here a convex, cylindrical, cup-shaped
or toroidal set in three dimensions, and a closed convex curve in
two dimensions. We require convexity for tubular surfaces, motivated
by the near-circular cross section generally observed for stable vortices.

\begin{figure}
\centering 
\includegraphics[width=0.6\textwidth]{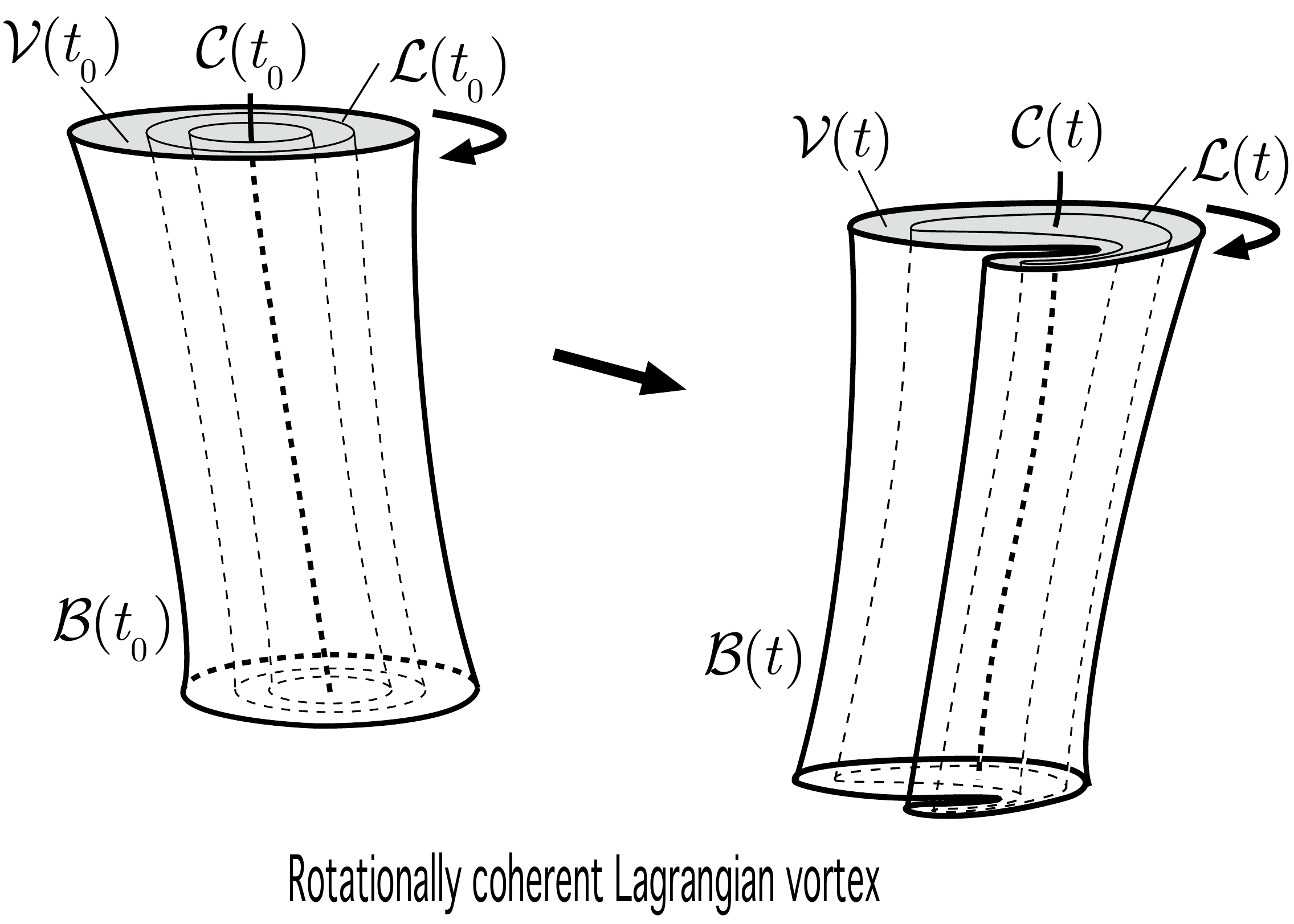}
\caption{Initial and current positions of a rotationally coherent Lagrangian
vortex $\mathcal{V}(t)$; a Lagrangian vortex boundary $\mathcal{B}(t);$
and a Lagrangian vortex center $\mathcal{C}(t)$. Also shown within
$\mathcal{V}(t)$ is a rotational LCS $\mathcal{L}(t)$, i.e., a material
surface along which volume elements exhibit the same intrinsic dynamic
rotation over the time interval $[t_{0},t_{1}].$ As a consequence,
$\mathcal{V}(t)$ experiences only tangential material filamentation
without global breakaway.}
\label{fig:rotational.LCS} 
\end{figure}

If the gradient $\partial_{x_{0}}\mathrm{LAVD}_{t_{0}}^{t_{1}}(x_{0})$
is nonzero along a tubular LAVD level surface, then this level surface
is surrounded by a continuous, nested family of tubular level surfaces
(Milnor 1963). The singular center of such a nested sequence of tubes,
with inward increasing LAVD values, gives a definition of a Lagrangian
vortex center. Similarly, the largest convex member of such a nested
tube family defines the boundary of a Lagrangian vortex. We summarize
these concepts in the following definition, with its geometry illustrated
in Fig. \ref{fig:rotational.LCS}.

\begin{defn} Over the finite time interval $[t_{0},t_{1}]$:

(i) A \emph{rotationally coherent Lagrangian vortex} is an evolving
material domain $\mathcal{V}(t)$ such that $\mathcal{V}(t_{0})$
is filled with a nested family of tubular level surfaces of $\mathrm{LAVD}_{t_{0}}^{t_{1}}(x_{0})$
with outward-decreasing LAVD values.

(ii) \emph{The boundary} $\mathcal{B}(t)$ of $\mathcal{V}(t)$ is
a material surface such that $\mathcal{B}(t_{0})$ is the outermost
tubular level surface of $\mathrm{LAVD}_{t_{0}}^{t_{1}}(x_{0})$ in
$\mathcal{V}(t_{0})$.

(iii) \emph{The center $\mathcal{C}(t)$ }of $\mathcal{V}(t)$\emph{
}is a material set $\mathcal{C}(t)$ such that $\mathcal{C}(t_{0})$
is the innermost member (maximum) of the $\mathrm{LAVD}_{t_{0}}^{t_{1}}(x_{0})$
level-surface family in $\mathcal{V}(t_{0})$. \end{defn} We refer
to the evolving positions $\mathcal{L}(t)$ of the tubular level sets
\[
\mathcal{L}(t_{0})=\left\{ x_{0}\in U(t_{0})\,:\,\mathrm{LAVD}_{t_{0}}^{t_{1}}(x_{0})=const.\right\} 
\]
as a rotational Lagrangian Coherent Structure \emph{(rotational LCS)},
as indicated in Fig. \ref{fig:rotational.LCS}. These LCSs give a
foliation of the evolving Lagrangian vortex $\mathcal{V}(t)$ into
tubes along which material elements complete the same intrinsic dynamic
rotation $\psi_{t_{0}}^{t}(x_{0})$.

Rotational LCSs, as well as Lagrangian vortices, their boundaries
and centers are material objects by definition. Therefore, their time
$t$ position is uniquely determined by Lagrangian advection of their
initial positions: 
\[
\mathcal{L}(t)=\mathcal{F}_{t_{0}}^{t}\left(\mathcal{L}(t_{0})\right),\qquad\mathcal{B}(t)=\mathcal{F}_{t_{0}}^{t}\left(\mathcal{B}(t_{0})\right),\qquad\mathcal{C}(t)=\mathcal{F}_{t_{0}}^{t}\left(\mathcal{C}(t_{0})\right),\qquad t\in[t_{0},t_{1}].
\]
Recent stretching-based definitions of Lagrangian vortices allow for
no filamentation in boundary of the vortex (Haller \& Beron--Vera
2013, Blazevski \& Haller 2014). In contrast, the LAVD-based definition
of a rotational LCS allows for tangential material filamentation.
The filamented part of the material surface, however, still rotates
together with the LCS without global breakaway.

The above definitions capture Lagrangian vortices with the simplest
(i.e., convex) geometry at time $t_{0}$. More generally, one may
allow for small tangential filamentation to be a priori present in
the vortex boundary even at time $t_{0}$. This involves the relaxation
of the convexity of $\mathcal{L}(t_{0})$ and $\mathcal{B}(t_{0})$
to material surfaces with small convexity deficiency, as discussed
along with other numerical aspects in Section \ref{sec:numerics}.

In geophysical flows over a rotating planet, rotationally coherent
Lagrangian vortices can directly be computed from the flow induced
in the curvilinear coordinate space instead of the curved surface
of the planet (cf. Appendix B). By construction, the resulting vortices
and their centers are invariant with respect to time-dependent rotations
and translations within the space of curvilinear coordinates.\footnote{Frame-invariance cannot be defined restricted to a curvilinear surface,
as Euclidean frame changes take the observer off the surface.} With this approach, one simply computes the classic Euclidean vorticity
of the longitudinal and latitudinal coordinate speeds, as opposed
to computing the vorticity in curvilinear coordinates.

\section{Rotationally coherent Eulerian vortices}

Over a short time interval $[t_{0},t+s]$ with $\left|s\right|\ll1$,
we can Taylor expand the LAVD field as 
\begin{equation}
\mathrm{LAVD}_{t_{0}}^{t+s}(x_{0})=\mathrm{LAVD}_{t_{0}}^{t}(x_{0})+\mathrm{IVD}(x(t;x_{0}),t)\cdot s+\mathcal{O}\left(s^{2}\right),\label{eq:Taylor_LAVD}
\end{equation}
with the \emph{instantaneous vorticity deviation }(IVD) defined as
\begin{equation}
\mathrm{IVD}(x,t):=\left|\omega(x,t)-\bar{\omega}(t)\right|.\label{eq:IVD}
\end{equation}
By the calculation \eqref{eq:objective}, the IVD field is objective.
By equation \eqref{eq:Taylor_LAVD}, $\mathrm{IVD}(x(t;x_{0}),t)$
describes the rate of change of the LAVD field at an initial condition
$x_{0}$ under increasing integration time.

\begin{figure}
\centering 
\includegraphics[width=0.3\textwidth]{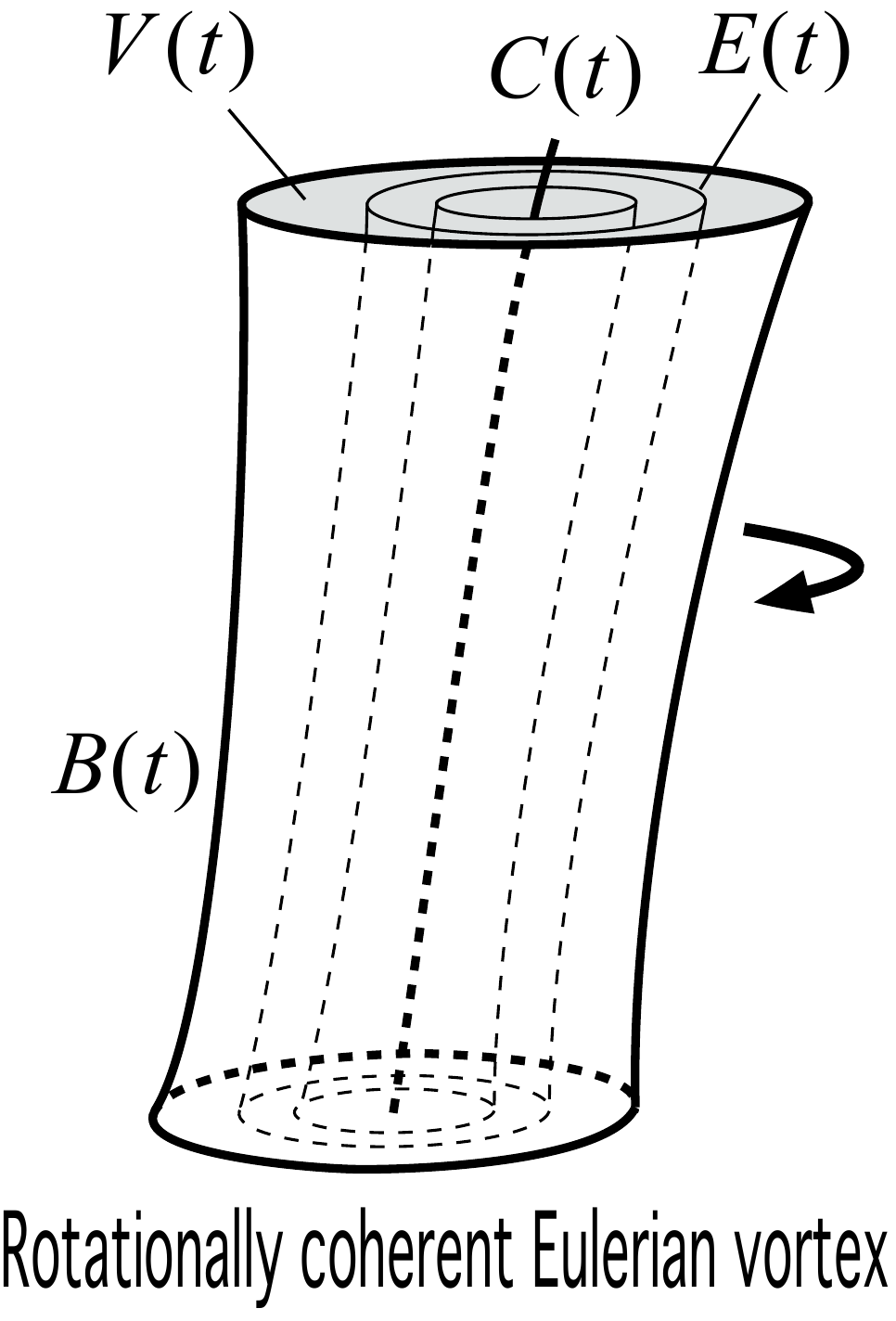}
\caption{A rotationally coherent Eulerian vortex $V(t)$, with boundary $B(t)$
and vortex center $C(t)$. Also shown within $V(t)$ is a rotational
ECS $E(t)$, i.e., a surface along which volume elements exhibit the
same intrinsic dynamic rotation rate instantaneously at time $t$.
No material coherence is guaranteed for the Eulerian vortex $V(t)$
under passive advection.}
\label{fig:rotational.ECS} 
\end{figure}

Using the IVD, we now introduce the instantaneous notion of a rotationally
coherent Eulerian vortex by taking the limit $t_{0},t_{1}\to t$ in
Definition 1. At a time $t\in[t{}_{0},t_{1}],$ such an Eulerian vortex
is composed of tubular surfaces along which the intrinsic rotation
rates $\dot{\psi}_{t}^{t}$ of fluid elements are equal. Indeed, by
formula \eqref{eq:psidef}, we have $\dot{\psi}_{t}^{t}(x)=\frac{1}{2}\mathrm{IVD}(x,t)=const.$
along these tubular surfaces. The following definition summarizes
the details for this objective Eulerian vortex concept. \begin{defn}
At a time instance $t\in[t_{0},t_{1}]:$

(i) A \emph{rotationally coherent Eulerian vortex} is a set $V(t)$
filled with a nested family of tubular level sets of $IVD(x,t)$ with
outwards non-increasing IVD values.

(ii) \emph{The boundary} $B(t)$ of $V(t)$ is the outermost level
surface of $\mathrm{IVD}(x,t)$ in $V(t)$.

(iii) \emph{The center $C(t)$ }of $V(t)$ is the innermost member
(maximum) of the $\mathrm{IVD}(x,t)$ level-surface family in $V(t)$.
\end{defn} A rotational Eulerian Coherent Structure (\emph{rotational
ECS}) is then just a level surface 
\begin{equation}
E(t)=\left\{ x\in U(t)\,:\,\mathrm{IVD}(x,t)=C_{0}(t)\right\} \label{eq:ECS}
\end{equation}
along which material elements experience the same dynamic rotation
rate $\dot{\psi}_{t}^{t}(x)=C_{0}(t)/2$. Unlike rotational LCSs,
individual rotational ECSs are instantaneous quantitates without a
well-defined evolution, unless $C_{0}(t)$ is specifically selected
as constant over time. We illustrate the geometry of Definition 2
in Fig. \ref{fig:rotational.ECS}.

\begin{figure}
\centering 
\includegraphics[width=0.5\textwidth]{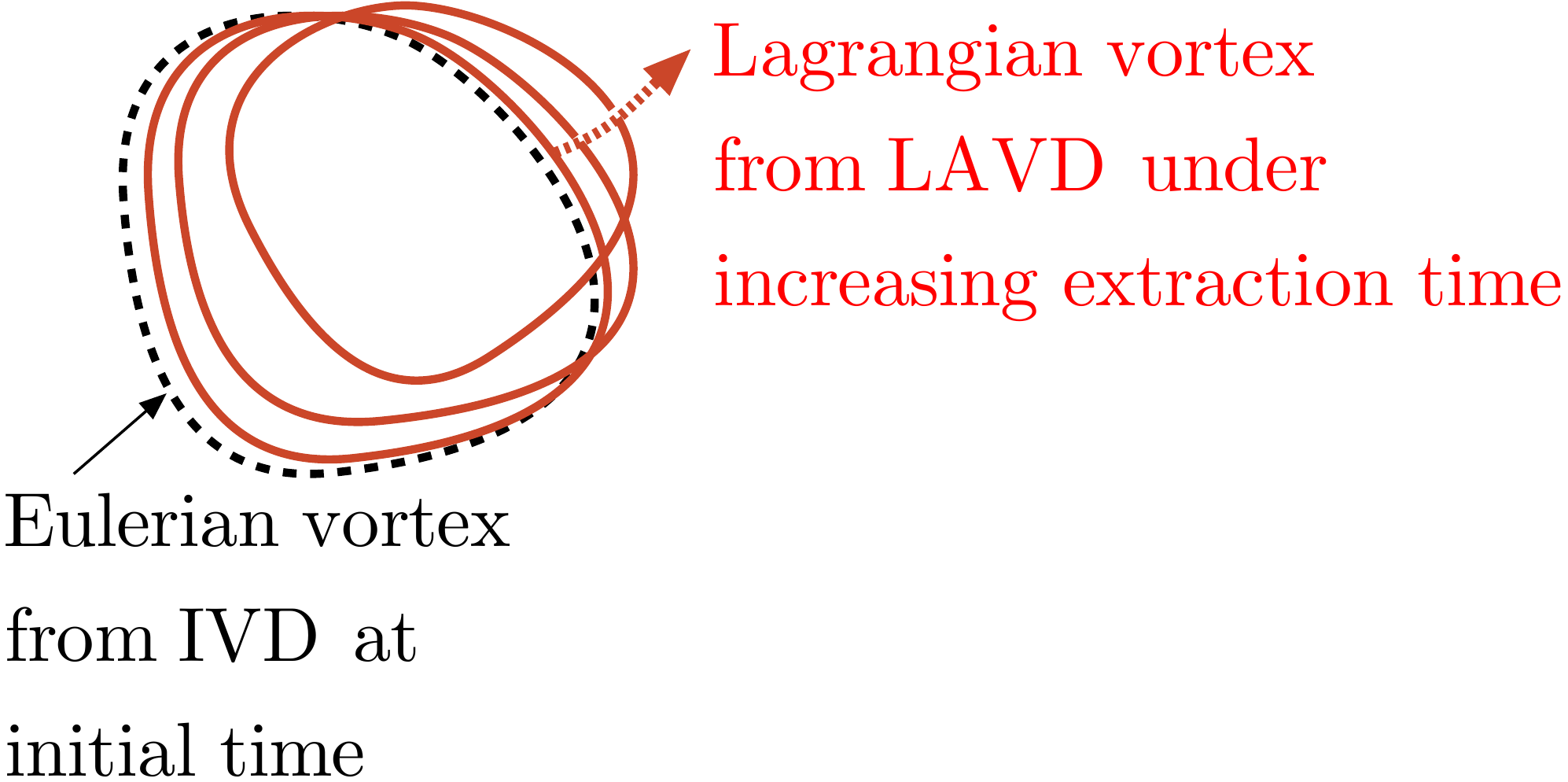}
\caption{A rotationally coherent Eulerian vortex is the limit of a rotationally
coherent Lagrangian vortex under vanishing extraction time (i..e,
integration) time.}
\label{fig:IVD-LAVD} 
\end{figure}

Given that\emph{ 
\[
\frac{d}{dt}\mathrm{LAVD}_{t_{0}}^{t}(x_{0})=\mathrm{IVD}(x(t;x_{0}),t),
\]
rotationally coherent Eulerian vortices are effectively the derivatives
of rotationally coherent Lagrangian vortices with respect to the length
of the extraction time of the latter.} Setting $t=t_{0}$ in \eqref{eq:Taylor_LAVD}
and observing that $\mathrm{LAVD}_{t_{0}}^{t_{0}}(x_{0})=0,$ we conclude
that the $\mathcal{B}(t_{0})$ initial positions of rotationally coherent
Lagrangian vortex boundaries evolve precisely from their Eulerian
coherent counterparts $B(t_{0})$ as the Lagrangian extraction time
$t_{1}-t_{0}$ increases from zero (see Fig. \ref{fig:IVD-LAVD}).

Physically, these Eulerian vortices are built of tubular surfaces
showing instantaneous coherence in the rate of their bulk material
rotation. This instantaneous coherence in rotation rates generally
does not imply sustained coherence in the finite rotation of material
trajectories released from these surfaces. Furthermore, the interior
of an evolving vortex $V(t)$ is not a material domain.

By the objectivity of the IVD, Definition 2 nevertheless gives an
objective definition of an Eulerian vortex, its center and boundary.
To our knowledge, no other objective, three-dimensional Eulerian vortex
definition has been proposed in the literature. Specifically, none
of the Eulerian criteria reviewed or proposed in Jeung \& Hussein
(1995), Haller (2005), and Chakraborty, Balachandar \& Adrian (2005)
are invariant under time-dependent rotations and translations of the
observer. Since truly unsteady flows have no distinguished frames
of reference (Lugt 1979), non-objective Eulerian vortex criteria do
not yield well-defined material vortices for unsteady flows.

A relevant discussion can be found in Jeung \& Hussein (1995) about
the \emph{$\left|\omega\right|$-criterion}, by which $\left|\omega\right|$
must exceed a preselect threshold within a vortex. This approach is
found intuitive but inadequate in Jeung \& Hussein (1995) for several
reasons. We agree with this general assessment, because the $\left|\omega\right|$-criterion
is threshold-dependent and not objective. In contrast, our rotationally
coherent Eulerian vortex definition in Definition 2 is based on a
threshold-independent and objective assessment of the level surface
topology of $\left|\omega-\bar{\omega}\right|.$

Importantly, for flows with zero mean vorticity in their frame of
definition, outermost convex tubular level sets of the vorticity magnitude
$\left|\omega\right|$ or of the enstrophy $\left|\omega\right|^{2}$
coincide with rotationally coherent Eulerian vortices. Therefore,
when properly interrogated, the vorticity and enstrophy distribution
of zero-mean-vorticity flows do reveal objective structures that can
be viewed as derivatives of rotationally coherent Lagrangian vortices.

\section{Rotationally coherent vortices in planar flows}

In a flow defined in the $(x_{1},x_{2})$ plane, the LAVD \eqref{eq:LAVD}
takes the simple form 
\begin{equation}
\mathrm{LAVD}_{t_{0}}^{t}(x_{0})=\intop_{t_{0}}^{t}\left|\omega_{3}(x(s;x_{0}),s)-\bar{\omega}_{3}(s)\right|\,\,ds,\label{eq:LAVD2D}
\end{equation}
with $\omega_{3}$ referring here to the $x_{3}$ component of the
vorticity vector $\omega(x,t)=(0,0,\omega_{3}(x,t))$, and with the
mean vorticity  $\bar{\omega}_{3}(s)$ 

By Definition 1, a rotational LCS evolves over the time interval $[t_{0},t_{1}]$
via advection from a closed and convex level curve of the $\mathrm{LAVD}_{t_{0}}^{t_{1}}(x_{0})$
computed in \eqref{eq:LAVD2D}. By the same definition, Lagrangian
vortex boundaries are outermost members of such rotational LCS families.
Similarly, Lagrangian vortex centers are advected positions of isolated
maxima of $\mathrm{LAVD}_{t_{0}}^{t_{1}}(x_{0})$.

By Definition 2, a rotational ECS at time $t$ is a closed and convex
level curve of 
\[
\mathrm{IVD}(x,t)=\left|\omega_{3}(x,t)-\bar{\omega}_{3}(t)\right|,
\]
around one of its local maxima. Accordingly, rotationally coherent
Eulerian vortices are outermost members of such nested curve families
with outwards non-increasing instantaneous IVD values.

For two-dimensional flows only, Haller (2016) shows that the \emph{relative
dynamic rotation} angle 
\[
\phi_{t_{0}}^{t}(x_{0}):=\frac{1}{2}\int_{t_{0}}^{t}\left[\omega_{3}(x(s;x_{0}),s)-\bar{\omega}_{3}(s)\right]\,ds
\]
is also an objective and dynamically consistent measure of rotation.
It measures the net rotation angle generated by the relative rotation
tensor around the $x_{3}$ axis, with sign changes in the rotation
accounted for.\footnote{In contrast, $\psi_{t_{0}}^{t}(x_{0})$ measures rotation about the
evolving instantaneous axis of rotation of $\Phi_{t_{0}}^{t}$, which
points in the $-x_{3}$ direction when $\omega_{3}-\bar{\omega}_{3}$
is negative.}

While $\psi_{t_{0}}^{t_{1}}(x_{0})$, as the total angle swept by
the relative rotation tensor over the time interval $[t_{0},t_{1}]$,
is always positive, the sign of the angle $\phi_{t_{0}}^{t_{1}}(x_{0})$
is unrestricted. In vortical regions preserving the pointwise the
sign of the Lagrangian vorticity $\omega_{3}(x(s;x_{0}),s)-\bar{\omega}_{3}(s)$,
contours of $\phi_{t_{0}}^{t_{1}}(x_{0})$, contours of $\mathrm{LAVD}_{t_{0}}^{t_{1}}(x_{0})$,
and contours of the \emph{Lagrangian-averaged vorticity (LAV)} 
\[
\mathrm{LAV}_{t_{0}}^{t_{1}}(x_{0}):=\int_{t_{0}}^{t_{1}}\omega_{3}(x(s;x_{0}),s)\,ds
\]
all coincide with each other, albeit generally correspond to different
values of the underlying scalar fields.\footnote{Note that $\mathrm{LAV}_{t_{0}}^{t_{1}}(x_{0})$ is not an objective
quantity, but its level curves are objectively defined.} In such regions, the sign of $\phi_{t_{0}}^{t_{1}}(x_{0})$ will
also carry objective information about the direction of the relative
rotation.

Differences in the contours of $\mathrm{LAVD}_{t_{0}}^{t_{1}}(x_{0})$,
$\phi_{t_{0}}^{t_{1}}(x_{0})$ and $\mathrm{LAV}_{t_{0}}^{t_{1}}(x_{0})$
will arise in regions where the sign of $\omega_{3}(x(s),s)-\bar{\omega}_{3}(s)$
crosses zero, i.e., near the boundaries of regions with a well-defined
sign in their deviation from the mean vorticity of the flow. In such
regions, connected level curves of $\mathrm{LAV}_{t_{0}}^{t_{1}}(x_{0})$
can group together initial conditions with substantially different
global rotational histories, as long as their final net rotations
are equal. The use of $\left.\mathrm{LAVD}\right._{t_{0}}^{t_{1}}(x_{0})$
eliminates such coincidental agreement in the net rotation angles.

\section{Geostrophic Lagrangian vortex centers are attractors for inertial
particles }

Consider a small spherical particle of radius $r_{0}$ and density
$\rho_{part}$ in a geostrophic flow of density $\rho$ and viscosity
$\nu$. Under the $\beta$-plane approximation, let $f$ denote the
Coriolis parameter (twice the local vertical component of the angular
velocity of the earth). Applying a slow-manifold reduction to the
Maxey--Riley equations (Maxey \& Riley 1983) in the limit of small
Rossby numbers, Beron--Vera et al. (2015) showed that the inertial
particle motion satisfies 
\begin{equation}
\dot{x}=v(x,t)+\tau(\delta-1)fJv(x,t)+\mathcal{O}(\tau^{2}),\qquad J=\left(\begin{array}{cc}
0 & -1\\
1 & 0
\end{array}\right),\label{eq:parteq}
\end{equation}
where 
\[
\delta=\frac{\rho}{\rho_{part}},\qquad\tau:=\frac{2r_{0}^{2}}{9\nu\delta}.
\]
Provenzale (1999) considered the Maxey--Riley equation in the same
physical setting, but without a slow-manifold reduction. His second-order
differential equation also included additional terms that either vanish
along the $\beta$-plane or appear at higher order in the reduced
first-order equation \eqref{eq:parteq} (cf. Beron--Vera et al. 2015
for more detail.)

Remarkably, in the limit of vanishing Rossby numbers, cyclonic attractors
for light particles $(\delta>1)$ and anticyclonic attractors for
heavy particles $(\delta<1)$ in eq. \eqref{eq:parteq} turn out to
be precisely the rotationally coherent vortex centers defined in Definition
1. The same Lagrangian vortex centers act as cyclonic repellers for
heavy particles and anticyclonic repellers for light particles. We
state these results in more detail as follows: \begin{thm} Assume
that $x_{0}^{*}\in U(t_{0})$ is the initial position of a rotationally
coherent Lagrangian vortex center whose relative rotation keeps constant
sign, i.e., 
\begin{equation}
\mathrm{sign}\left[\omega_{3}(x_{0}^{*},t)-\bar{\omega}_{3}(t)\right]=\mu(x_{0}^{*}),\qquad t\in[t_{0},t_{1}],\label{eq:cond1}
\end{equation}
for an appropriate sign constant $\mu(x_{0}^{*})\in\{-1+1\}$. Then,
for $\tau>0$ small enough, the following hold:

(i) In a cyclonic ($\mu(x_{0}^{*})f>0$) rotationally coherent vortex,
there exists a finite-time attractor (repeller) for light (heavy)
particles that stays $\mathcal{O}(\tau)$ close to the vortex center
$\mathcal{C}(t)$.

(ii) In an anticyclonic ($\mu(x_{0}^{*})f<0$) rotationally coherent
vortex, there exists a finite-time attractor (repeller) for heavy
(light) particles that stays $\mathcal{O}(\tau)$ close to the vortex
center $\mathcal{C}(t)$. \end{thm} \emph{Proof:} See the Appendix
C.

\begin{figure}
\centering \includegraphics[width=0.6\textwidth]{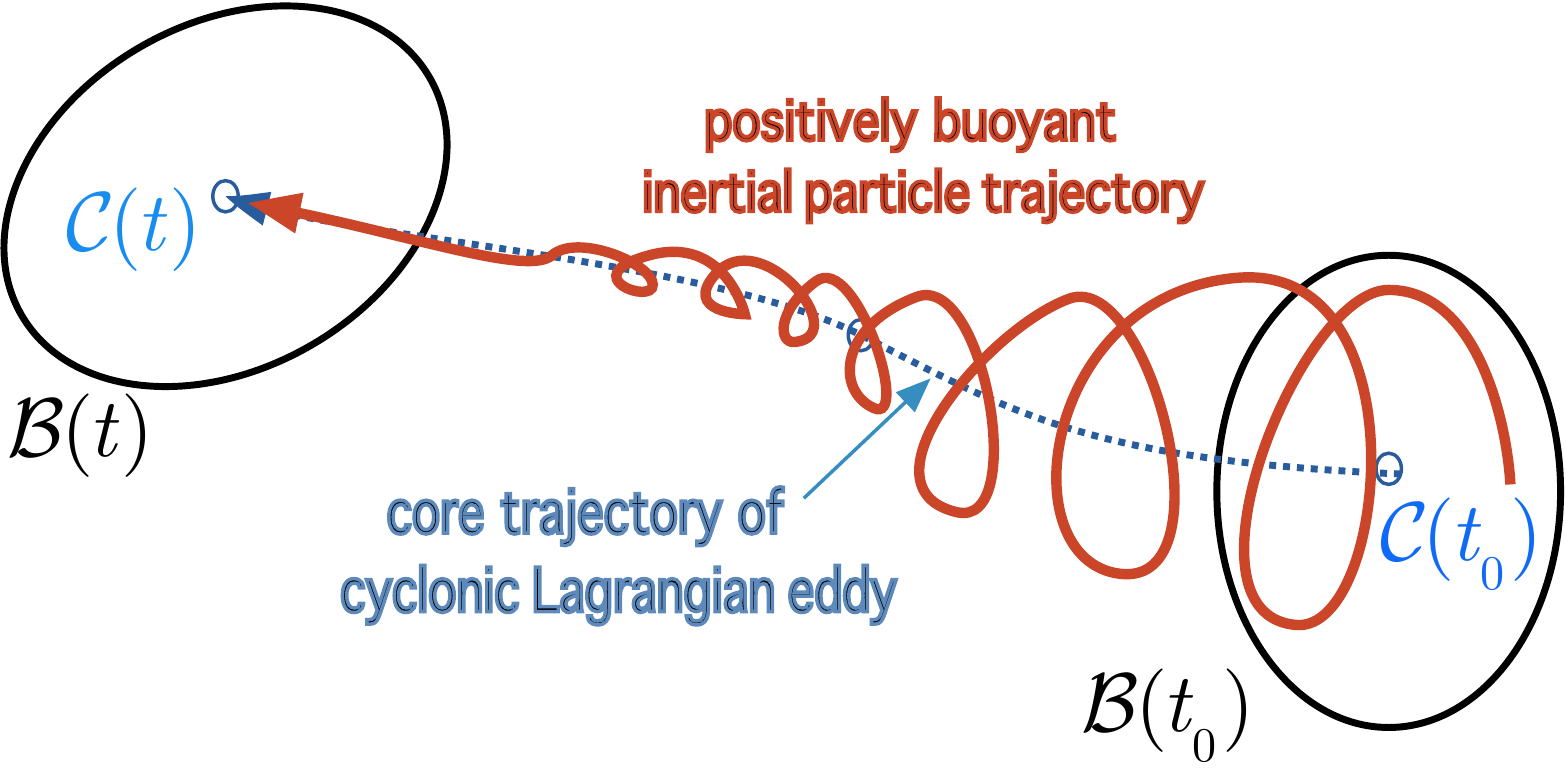}
\caption{By Theorem 2, cyclonic Lagrangian eddy centers evolving from LAVD
maxima act as observed attractors for small, positively buoyant particles,
such as drifters and floating debris.}
\label{fig:inertial.attractor} 
\end{figure}

Theorem 2 provides an independent, experimentally verifiable justification
for defining vortex centers as in Definition 1 for geostrophic flows.
Specifically, positively buoyant drifters or floating debris released
well inside a cyclonic oceanic eddy will spiral onto the evolving
Lagrangian vortex center identified from Definition 1 (Fig. \ref{fig:inertial.attractor}).
We will illustrate this effect using simulated inertial particle motion
on satellite-based ocean velocities in Section \ref{sub:Two-dimensional-geostrophic-flow}.

\section{Numerical aspects}\label{sec:numerics}

Computing a rotation angle from the classic polar decomposition requires
the computation of the deformation gradient $F_{t_{0}}^{t}$ (cf.
\eqref{eq:polardecomp} in Appendix A). This is either achieved by
the numerical differentiation of fluid trajectories with respect to
their initial conditions, or by solving the equation of variations
$\frac{d}{dt}F_{t_{0}}^{t}=\nabla v(x(t;t_{0},x_{0}),t)F_{t_{0}}^{t}$,
whose solutions typically grow exponentially. Either way, computing
polar rotation has the same long-time numerical sensitivity that arises
in computing the invariants of the Cauchy-Green strain tensor.

In contrast, computing rotationally coherent vortices based on Definition
1 only requires the integration of the normed vorticity deviation
along fluid trajectories. This is the simplest to do simultaneously
with trajectory integration, solving the extended system of differential
equations 
\begin{eqnarray}
\frac{dx}{dt} & = & v(x,t),\nonumber \\
\frac{d}{dt}\mathrm{LAVD}_{t_{0}}^{t} & = & \left|\nabla\times v(x,t)-\overline{\nabla\times v}(t)\right|,\label{eq:combined_ODE}
\end{eqnarray}
over a grid of initial conditions over the time interval $[t_{0},t_{1}].$
In our experience, however, features of the $\mathrm{LAVD}_{t_{0}}^{t}(x_{0})$
turn out to be sharper when the trajectory ODE $\dot{x}=v(x,t)$ is
solved first, and the vorticity is subsequently integrated along trajectories.
This is because adaptive ODE solvers make different decisions about
time steps when the trajectory ODE is amended with the ODE for the
LAVD field. This is especially so when both ODEs are solved simultaneously
over large grids of initial conditions.

The computational domain for solving \eqref{eq:combined_ODE} is just
$U(t)\equiv U$ in case of a closed flow with a fixed boundary. In
open flows, the focus may be on vortices on a smaller domain. In that
case, the domain should still be chosen large enough so that the averaged
vorticity $\overline{\nabla\times v}(t)$ is representative of the
overall mean rotation of the fluid mass under study. In geophysical
flows, this mean rotation is expected to be zero, which is confirmed
by our calculations even for domains of the size of a few degrees.

For two-dimensional flows, we first identify local maxima of the LAVD
field, then extract nearby closed LAVD level curves. In all our computations,
we use the level set function of MATLAB for this purpose, and identify
the closedness of a level curve by probing the output from this function.
Definition 1 then requires the identification of the outermost convex
LAVD level curve around an LAVD maximum as vortex boundary. This convexity
requirement is conservative, ensuring that the material vortex starts
out unfilamented at the initial time $t_{0}$. At later times, our
approach allows for tangential filamentation in the advected LCS due
to local strain, but disallows large-scale filamentation arising from
differences in the bulk rotation along material filaments. As a consequence,
filaments developed by rotational LCSs rotate together with the main
body of the underlying material vortex.

In actual computations, one reason to relax strict convexity for closed
LAVD level surfaces is that they are numerically represented by discrete
polygons. The more vertices such a polygon has, the more likely it
is that the polygon is not convex, even if the approximated level
curve is. A second reason for relaxing convexity is to remove the
conceptual asymmetry of Fig. \ref{fig:rotational.LCS}, allowing for
small tangential filamentation even in the initial positions of vortex
boundaries. A third reason for relaxing strict convexity in multi-scale
data sets is the presence of smaller-scale vortices near the perimeter
of a larger-scale vortex. This necessitates the use of an appropriately
coarse-grained notion for the boundary of the larger vortex.

\begin{figure}
\centering 
\includegraphics[width=0.8\textwidth]{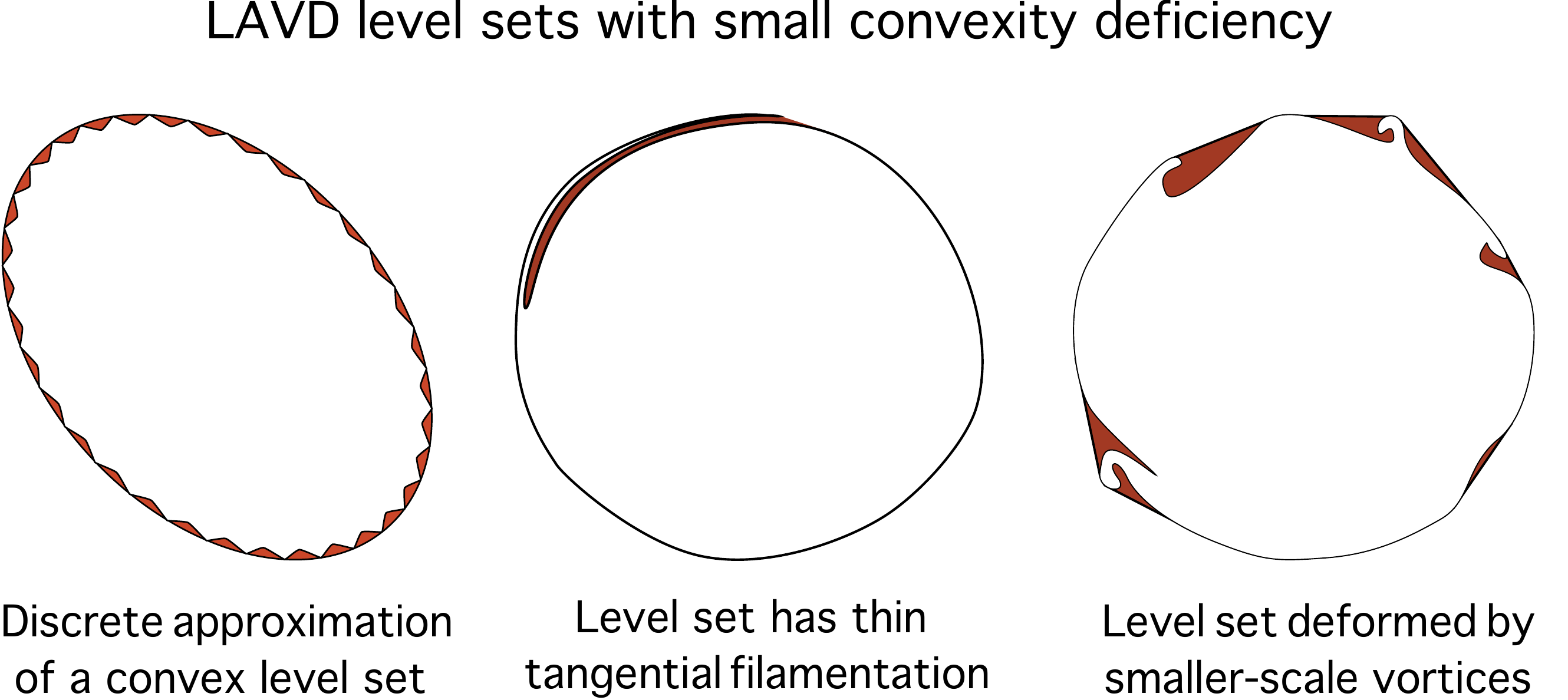}\caption{Cases of closed LAVD level sets benefiting from the relaxation of
convexity to small convexity deficiency. Shaded area indicates the
area difference between the level set and its convex hull. (1) A convex
LAVD level set with a non-convex approximating polygon arising from
discretization. (2) Closed LAVD level set with minor initial filamentation
in the tangential direction. (3) Closed LAVD level set that is only
convex after a filtering of smaller-scale vortices.}
\label{fig:convexity_deficiency} 
\end{figure}

In two dimensions, addressing the finite-grid, the initial tangential
filamentation, and the multi-scale challenges can be achieved by allowing
for a small convexity deficiency in the LAVD level curves (Gonzalez
\& Woods 2008, Batchelor \& Whealan 2012). Here, we define the \emph{convexity
deficiency} of a closed curve in the plane as the ratio of the area
difference between the curve and its convex hull to the area enclosed
by the curve. In Fig. \ref{fig:convexity_deficiency}, we show cases
of closed LAVD level sets with small convexity deficiency. In three
dimensions, we require small convexity deficiency for the one-dimensional
intersection of LAVD level surfaces with a family $\left\{ \mathcal{P}_{i}\right\}$
of planes transverse to the expected vortex center curve $\mathcal{C}(t_{0})$.

In our computations, we set the convexity deficiency bound to $10^{-3}$
or lower, which, we find to produce robust results for the first two
cases covered in Fig. \ref{fig:convexity_deficiency}. Applications
to multi-scale data sets (last panel of Fig. \ref{fig:convexity_deficiency})
will likely require a somewhat higher bound. In general, increasing
the convexity deficiency bound produces larger eddies that also tend
to filament more.

Small-scale local maxima of the LAVD function also arise due to numerical
or observational noise in the velocity data. To eliminate the resulting
artificial vortex candidates, we choose to ignore closed LAVD contours
whose arclength falls below a minimal threshold. In a given application,
this threshold should be chosen below the minimal vortex perimeter
that is expected to be reliably resolved by the data set.

We summarize the extraction algorithm for rotationally coherent Lagrangian
vortices in two- and three-dimensional flows in the tables entitled
Algorithm 1 and Algorithm 2 below. The extraction of their rotationally
coherent Eulerian counterparts follows the same steps (2)-(4), but
applied to the function $\mathrm{IVD}(x,t)$ instead of $\mathrm{LAVD}_{t_{0}}^{t}(x_{0}).$

\begin{algorithm}
\caption{Coherent Lagrangian vortex boundaries and vortex centers for 2-dimensional
flows}
\label{alg:algorithm1} \textbf{Input:} A 2-dimensional, time resolved
velocity field 
\begin{enumerate}
\item For a two-dimensional grid of initial conditions $x_{0}$, compute
the Lagrangian-Averaged Vorticity Deviation $\mathrm{LAVD}_{t_{0}}^{t_{1}}(x_{0})=\int\limits _{t_{0}}^{t_{1}}|\omega_{3}\left(x(s;x_{0}),s\right)-\bar{\omega}_{3}(s)|ds$. 
\item Detect initial positions $\mathcal{C}(t_{0})$ of vortex centers as
local maxima of $\mathrm{LAVD}_{t_{0}}^{t_{1}}(x_{0})$. 
\item Seek initial vortex boundaries $\mathcal{B}(t_{0})$ as outermost,
closed contours of $\mathrm{LAVD}_{t_{0}}^{t_{1}}(x_{0})$ satisfying
all the following:

\begin{enumerate}
\item $\mathcal{B}(t_{0})$ encircles a vortex center $\mathcal{C}(t_{0})$. 
\item $\mathcal{B}(t_{0})$ has arclength exceeding a threshold $l_{\min}$. 
\item $\mathcal{B}(t_{0})$ has convexity deficiency less than a bound $d_{\max}$. 
\end{enumerate}
\end{enumerate}
\textbf{Output: }Initial positions of rotationally coherent Lagrangian
vortex boundaries (closed curves) and vortex centers (isolated points)
with respect to the time interval $[t_{0},t_{1}].$ 
\end{algorithm}

\begin{algorithm}[h]
\caption{Coherent Lagrangian vortex boundaries and vortex centers for 3-dimensional
flows}
\label{alg:algorithm2} \textbf{Input:} A 3-dimensional, time-resolved
velocity field 
\begin{enumerate}
\item For a three-dimensional grid of initial conditions $x_{0}$, compute
the Lagrangian-Averaged Vorticity Deviation $\mathrm{LAVD}_{t_{0}}^{t_{1}}(x_{0})=\int\limits _{t_{0}}^{t_{1}}|\omega\left(x(s;x_{0}),s\right)-\bar{\omega}(s)|ds$. 
\item Detect initial positions $\mathcal{C}(t_{0})$ of vortex centers as
local maximum curves (singular level sets) of $\mathrm{LAVD}_{t_{0}}^{t_{1}}(x_{0})$.
This is best done by locating and connecting local maximum points
$\hat{\mathcal{C}}_{i}(t_{0})$ of $\mathrm{LAVD}_{t_{0}}^{t_{1}}(x_{0})$
in a family $\left\{ \mathcal{P}_{i}\right\} _{i=1}^{N}$ of $N$
parallel planes. 
\item Seek initial vortex boundaries $\mathcal{B}(t_{0})$ as outermost
tubular level surfaces of $\mathrm{LAVD}_{t_{0}}^{t_{1}}(x_{0})$
whose intersections $\mathcal{\hat{\mathcal{B}}}_{i}(t_{0})=\mathcal{B}(t_{0})\cap\mathcal{P}_{i}$
with the planes $\mathcal{P}_{i}$ satisfy the following for $i=1,\ldots,N$:

\begin{enumerate}
\item $\mathcal{\hat{\mathcal{B}}}_{i}(t_{0})$ encircles a vortex center
$\hat{\mathcal{C}}_{i}(t_{0})$. 
\item $\mathcal{\hat{\mathcal{B}}}_{i}(t_{0})$ has arclength exceeding
a threshold $l_{\min}$. 
\item $\mathcal{\hat{\mathcal{B}}}_{i}(t_{0})$ has convexity deficiency
less than a bound $d_{\max}$. 
\end{enumerate}
\end{enumerate}
\textbf{Output: }Initial positions of rotationally coherent Lagrangian
vortex boundaries (tubular surfaces) and vortex centers (curves) with
respect to the time interval $[t_{0},t_{1}].$ 
\end{algorithm}

The procedure outlined in Algorithm 2 is mathematically well-defined
by the level-surface topology of the LAVD. Step 2 of the algorithm
describes one possible numerical extraction scheme for these level
surfaces, using their intersections with planes transverse to the
anticipated vortex center $\mathcal{C}(t_{0})$. (A MATLAB implementation
of this algorithm is available under https://github.com/LCSETH.) Complex
flow geometries with multiple vortices and a priori unknown vortex
orientations will require more involved numerical approaches to level-set
extraction.

To implement Algorithms 1 and 2 in the forthcoming
examples, we use a variable-order Adams--Bashforth--Moulton solver
(ODE113 in MATLAB) for trajectory advection. The absolute and relative
tolerances of the ODE solver are chosen as $10^{-6}$ or higher. In
Sections 10.3, 10.4, 10.5 and 11.2, we use cubic and bilinear interpolation
schemes for computing pointwise vorticity values for the IVD and LAVD
functions, respectively. The lower-order interpolation suffices for
the LAVD calculation, because trajectory advection smoothes out the
LAVD contours in our experience.

\section{Two-dimensional examples}

\subsection{Planar Euler flows\label{sub:Planar-unsteady-Euler}}

On any solution of the two-dimensional Euler equation on the $(x_{1},x_{2})$
coordinate plane, the scalar vorticity value $\omega_{3}(x,t)$ is
preserved along trajectories. Therefore, the LAVD defined in \eqref{eq:LAVD}
simplifies to 
\[
\mathrm{LAVD}_{t_{0}}^{t}(x_{0})=(t-t_{0})\left|\omega_{3}(x_{0},t_{0})-\bar{\omega}_{3}(t_{0})\right|=(t-t_{0})\,\mathrm{IVD}(x_{0},t_{0}),
\]
Consequently, at time $t_{0}$, the boundaries and centers of all
rotationally coherent Lagrangian vortices coincide with those of rotationally
coherent Eulerian vortices. Specifically, ${\cal B}(t_{0})=B(t_{0})$
are outermost, closed and convex level curves of $\left|\omega_{3}(x_{0},t_{0})-\bar{\omega}_{3}(t_{0})\right|$,
encircling a local maximum ${\cal C}(t_{0})=C(t_{0})$ of $\left|\omega_{3}(x_{0},t_{0})-\bar{\omega}_{3}(t_{0})\right|$.

If the planar Euler flow is steady, then 
\[
\frac{d\omega(x(t))}{dt}=\nabla\omega(x(t))\cdot v(x(t))=0.
\]
Consequently, as long as the vorticity is not constant on open sets,
vorticity contours coincide with streamlines. In that case, both the
Lagrangian and the Eulerian vortices are bounded by outermost closed
streamlines, with their centers marked by a center-type fixed point,
as expected.

Despite these formal LAVD calculations for planar Euler flows, one
must remember: the eternal conservation of vorticity in these flows
creates a high-degree of degeneracy for the rotation of fluid parcels.
Indeed, despite the temporal and spatial complexity of a planar Euler
velocity field $v(x,t)$, a fluid element traveling along a fluid
trajectory $x(t)$ will keep its initial angular velocity (i..e, half
of its vorticity at $t_{0}$ ) for all times $t$. This property locks
the initial conditions of initially rotationally coherent fluid parcels
to the same LAVD level curve for all times, no matter how much these
parcels separate or deform in the meantime. Specifically, even if
material filaments break away transversely from a vortical region
and undergo high stretching and global filamentation, they will eternally
keep the exact same pointwise angular velocity they initially acquired
near the vortex (see, e.g., the Eulerian vortex interaction simulation
of Dritschel and Waugh 1992 for a striking example).

This rotational degeneracy of the planar Euler equation is lost under
the addition of the slightest viscous dissipation, compressibility
or three-dimensionality. Under any of these regularizing perturbations,
LAVD-based vortex detection can be applied to reveal the Lagrangian
signature of vortex interactions in the perturbed flow. We illustrate
this in a viscous perturbation of the contour dynamics simulation
of Dritschel and Waugh (1992) in Section 10.4.

\subsection{Irrotational vortices}

\label{Ex:irrot}

Although physically unrealizable, swirling flows with regions of zero
vorticity are important theoretical models. The classic irrotational
vortex flow is a given by the two-dimensional, circularly symmetric
velocity field

\begin{equation}
v(x)=\left(\begin{array}{c}
\frac{-\alpha x_{2}}{x_{1}^{2}+x_{2}^{2}}\\
\frac{\alpha x_{1}}{x_{1}^{2}+x_{2}^{2}}
\end{array}\right).\label{eq:irrot}
\end{equation}
The vorticity of this flow is identically zero, implying $\mathrm{LAVD}_{t_{0}}^{t}(x_{0})\equiv0$
and $IVD(x)\equiv0$ for any choice of $t_{0}$ and $t$. This may
seem to be at odds with the fact that all particles move on circular
orbits, and hence line elements tangent to the trajectories exhibit
one full rotation over the period of the circular orbit.

However, the tangents are special line elements and are not representative
of the overall bulk rotation of fluid elements. Material fibers transverse
to the trajectory tangents all rotate at different speeds. The average
rate of rotation for all line elements emanating from a given point
is zero, as was already pointed out by Helmholtz (1858) in a related
debate with Bertrand (1873) (cf. Truesdell \& Rajagopal 2009, Haller
2016). Therefore, an infinitesimal fluid volume has no experimentally
measurable bulk rigid-body rotation component in its deformation,
as indeed demonstrated by Shapiro (1961).

The irrotational vortex flow \eqref{eq:irrot} is composed of a single,
degenerate LAVD and IVD level set $\mathrm{LAVD}_{t_{0}}^{t}(x_{0})=\mathrm{IVD}(x)\equiv0$,
as opposed to a nested family of codimension-one tubular level sets
with outward decreasing LAVD and IVD values. Therefore, the flow \eqref{eq:irrot}
does not satisfy our Definitions 1 and 2, and hence does not qualify
as a rotationally coherent Lagrangian or Eulerian vortex. A similar
conclusion holds on the irrotational domains of the Rankine and the
Lamb--Oseen vortices (Majda \& Bertozzi 2002), or more generally,
in a recirculation region of any potential flow. We note that irrotational
vortices are also either explicitly excluded by most systematic vortex
criteria or fail the test for being a vortex according to these criteria.

\subsection{Direct numerical simulation of two-dimensional turbulence}

\label{Ex:2Dturb} We solve numerically the forced Navier--Stokes
equation 
\begin{equation}
\partial_{t}v+v\cdot\nabla v=-\nabla p+\nu\Delta v+f,\qquad\nabla\cdot v=0,\label{eq:NS}
\end{equation}
for a two-dimensional velocity field $v(x,t)$ with $x=(x_{1},x_{2})\in U=[0,2\pi]\times[0,2\pi]$.
We use a pseudo-spectral code with viscosity $\nu=10^{-5}$. We evolve
a random-in-phase velocity field in the absence of forcing ($f=0$)
until the flow is fully developed, then turn on a random-in-phase
forcing (cf. Farazmand et al. 2013). We identify this latter time
instance with the initial time $t_{0}=0$, and run the simulation
till the final time $t_{1}=50.$

\begin{figure}
\centering
\subfloat[]{\includegraphics[width=0.8\textwidth]{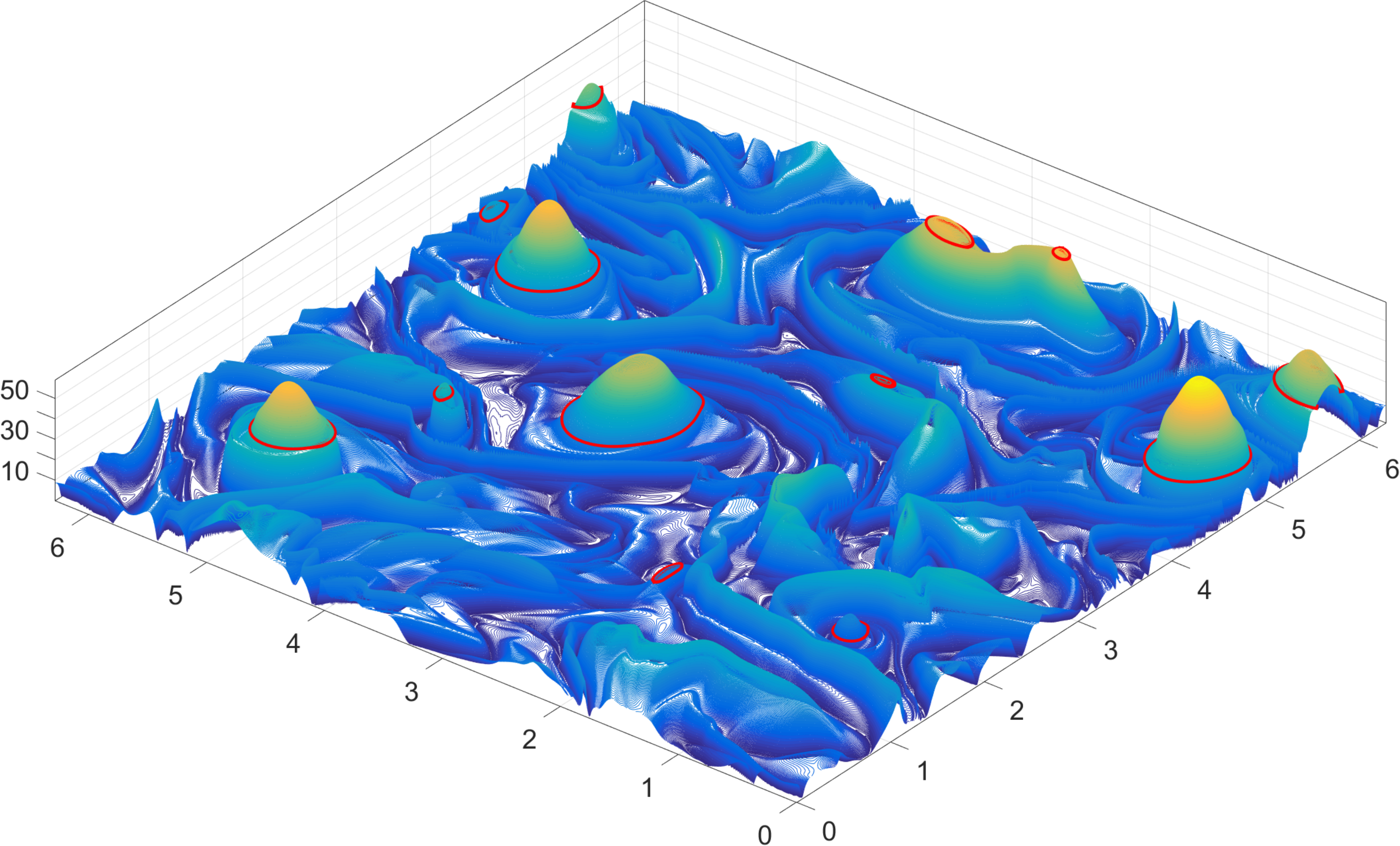}}\\
\subfloat[]{\includegraphics[width=0.4\textwidth]{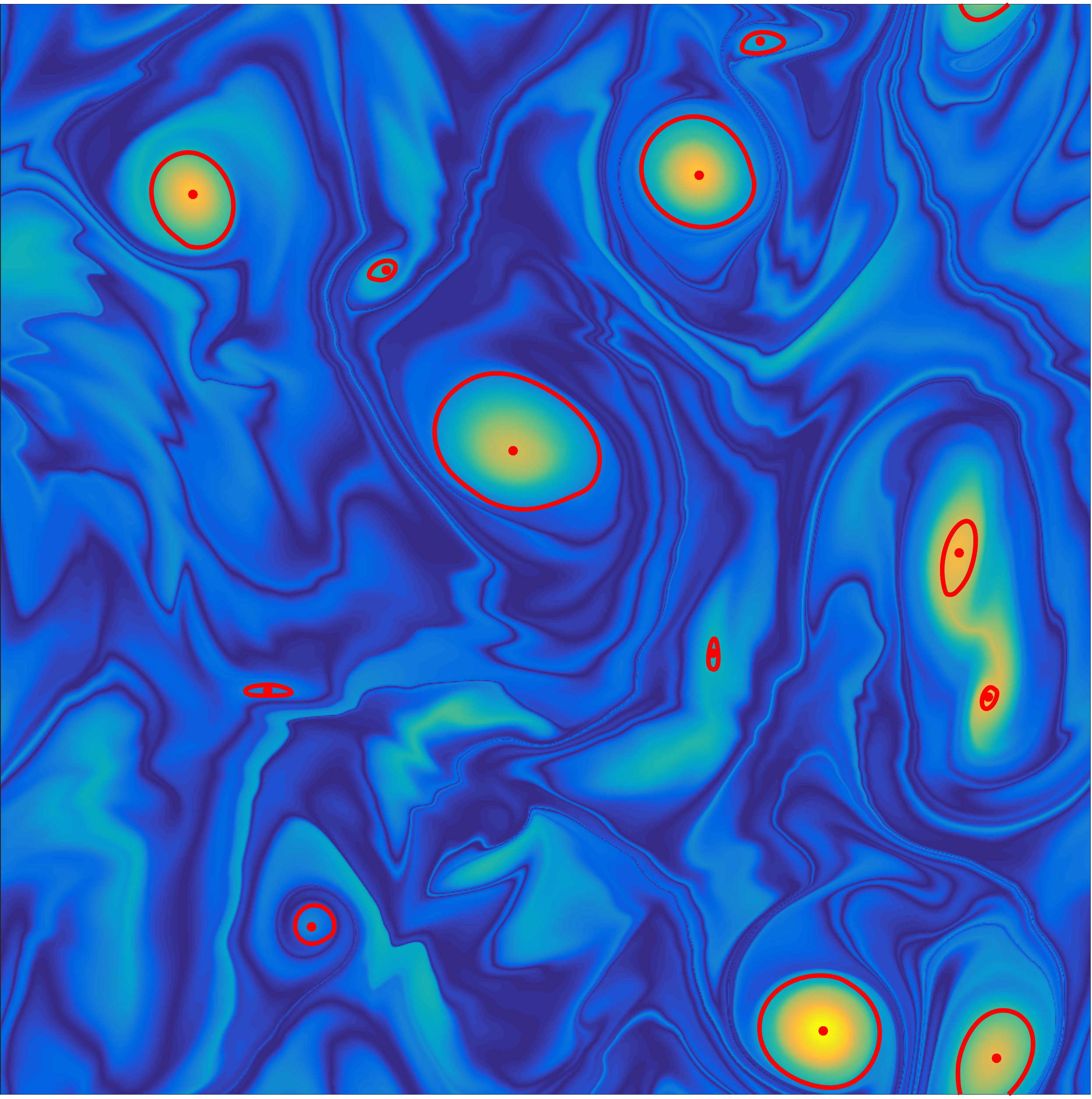}}\qquad
\subfloat[]{\includegraphics[width=0.4\textwidth]{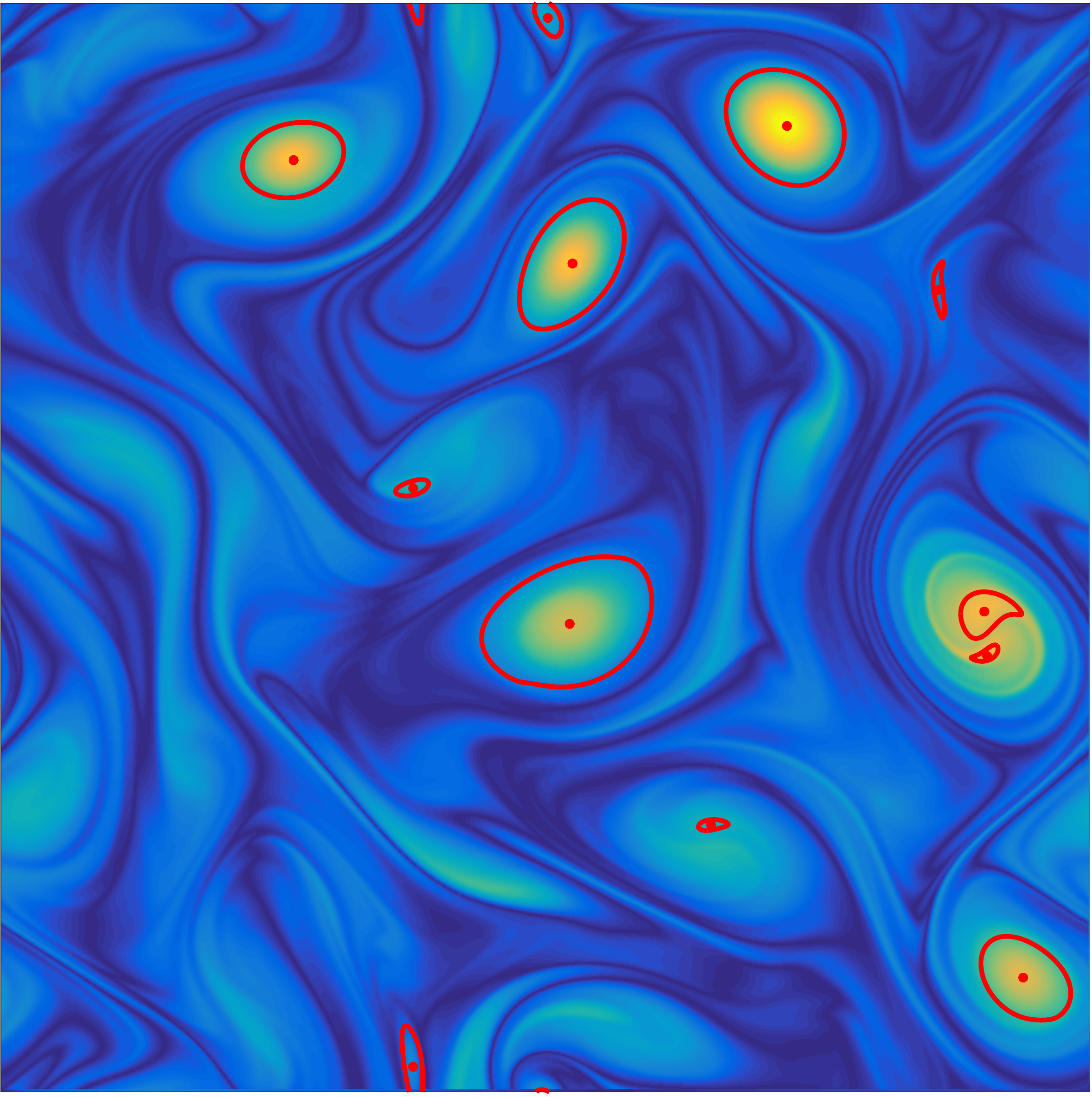}}\qquad
\caption{(a) Three-dimensional contour plot of $\mathrm{LAVD}_{0}^{50}(x_{0})$.
Lagrangian vortex boundaries (outermost rotational LCSs) extracted
from Algorithm 1 are shown in red. (b) Vortex boundaries shown in
the flow domain at initial time $t=0$. Shown in the background is
the contour plot of $\mathrm{LAVD}_{0}^{50}(x_{0})$ for reference.(c)
Advected Lagrangian vortex boundaries at the final time $t_{1}=50$.
(See the on-line supplemental movie M1 for the complete advection
sequence of the vortex boundaries).}
\label{fig:2Dturbulence} 
\end{figure}

To construct the LAVD field, we advect trajectories from an initial
grid of $1024\times1024$ points over the time interval $[0,50]$.
We integrated the vorticity deviation norm separately (as opposed
to solving the combined ODE \eqref{eq:combined_ODE}), using 1200
vorticity values along each trajectory, equally spaced in time. Figure
\ref{fig:2Dturbulence} shows the results superimposed on the contours
of $\mathrm{LAVD}_{0}^{50}(x_{0})$. In this computation, we have
set the minimum arc-length, $l_{\mathrm{min}}=0.3$ and convexity
deficiency bound $d_{\mathrm{max}}=10^{-4}$.

Figure \ref{fig:2Dturbulence} shows the Lagrangian vortex boundaries
extracted from Definition 1 using Algorithm 1 at the initial time
$t_{0}=0$. In Figure \ref{fig:2Dturbulence}, we confirm the Lagrangian
rotational coherence of these vortex boundaries by advecting them
to the final time $t_{1}=50$. As expected, the vortex boundaries
do not give in the general trend of exponential stretching and folding
observed for generic material lines. Instead, they display only local
(tangential) filamentation. The complete advection sequence over the
time interval $[0,50]$ is illustrated in the online supplemental
movie M1.

\begin{figure}
\centering 
\subfloat[]{\includegraphics[width=0.4\textwidth]{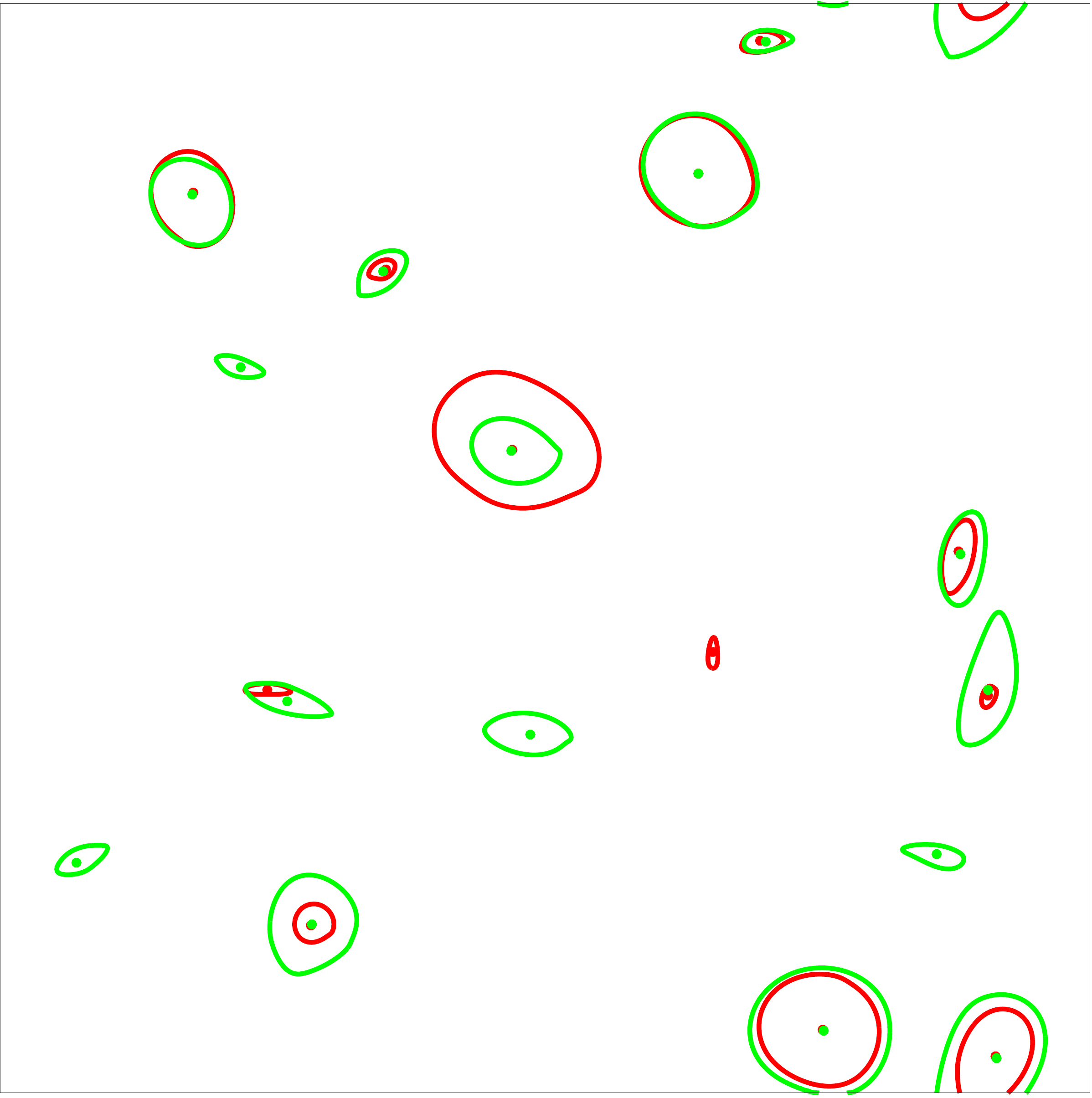}}\\
\subfloat[]{\includegraphics[width=0.4\textwidth]{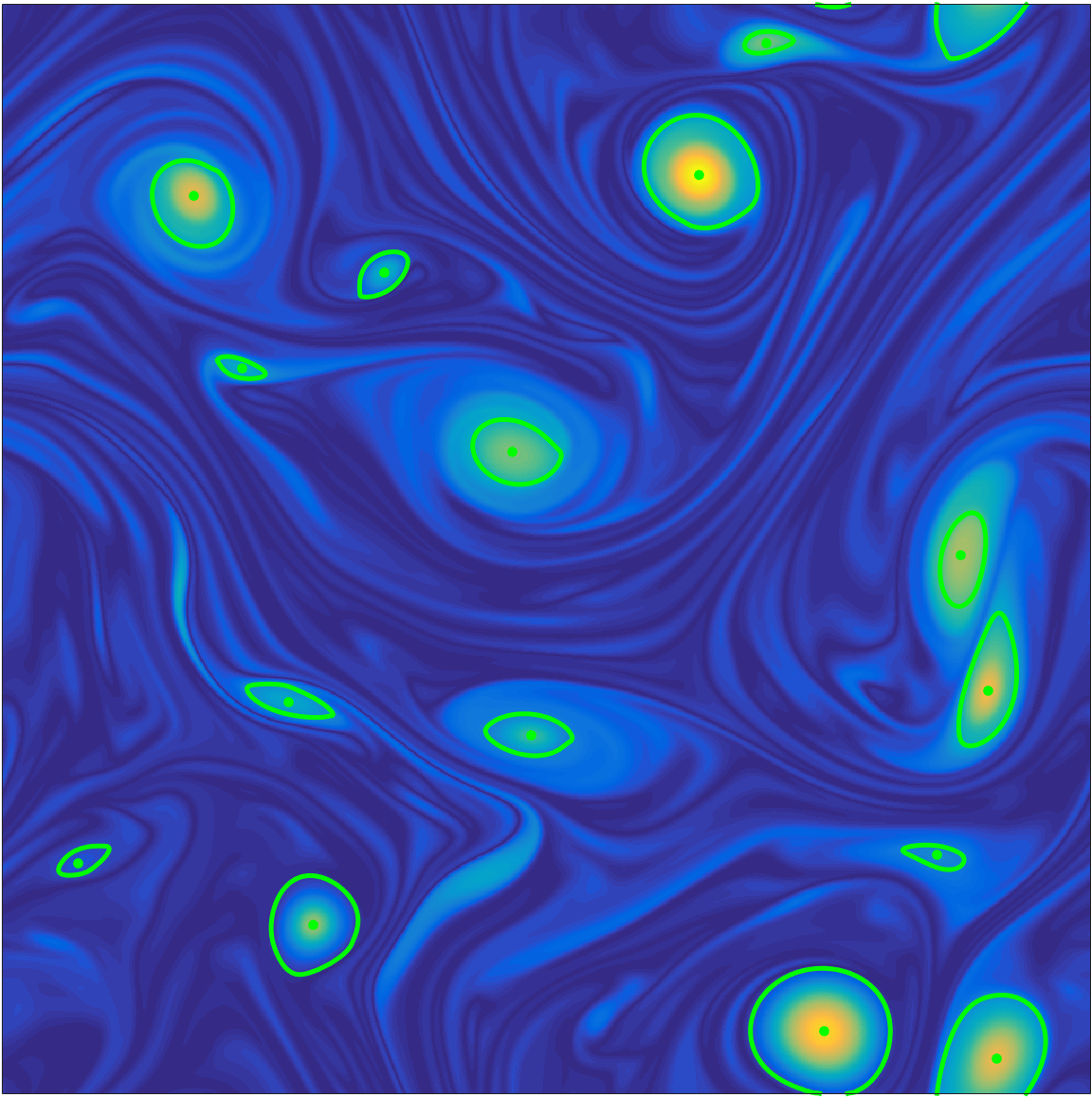}}\qquad
\subfloat[]{\includegraphics[width=0.4\textwidth]{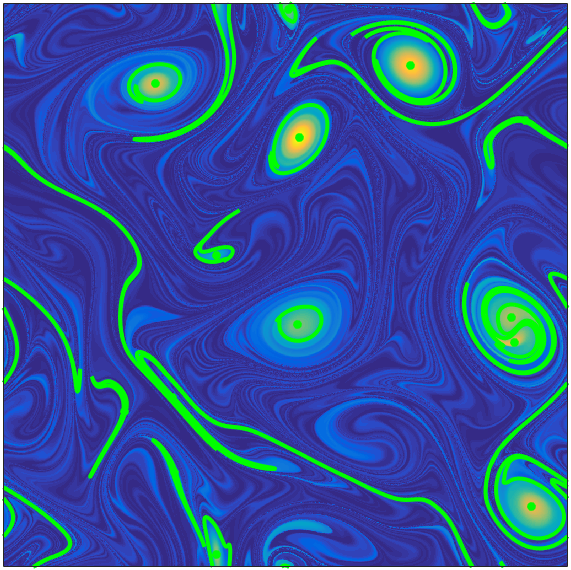}}
\caption{(a) Lagrangian (red) and Eulerian (green) rotationally coherent vortices
at time $t_{0}$ in the two-dimensional turbulence example. (b) Rotationally
coherent Eulerian vortex boundaries and centers at time $t_{0}$,
with the $\mathrm{IVD}(x,t_{0})$ field shown in the background. (c)
The same objects advected passively to the final time $t_{1}=50$,
with $\mathrm{IVD}(x,t_{1})$ shown in the background. }
\label{fig:2Dturbulence_IVD} 
\end{figure}

In Fig. \ref{fig:2Dturbulence_IVD}a, we show the time $t_{0}$ positions
of rotationally coherent Eulerian vortices in green, along with their
Lagrangian counterparts in red. Passively advected positions of the
Eulerian vortices are shown in Fig. \ref{fig:2Dturbulence_IVD}b,
displaying substantial material fingering into their surroundings.
While this lack of full material coherence over time is expected for
the Eulerian vortices, some of them are impressively close to their
Lagrangian counterparts at the initial time $t_{0}$, even though
they are extracted just from an instantaneous analysis. At the same
times, other Eulerian vortices without nearby Lagrangian counterparts
disintegrate completely under advection, as seen in Fig. \ref{fig:2Dturbulence_IVD}b.

\subsection{Interaction of two unequal vortices}

Motivated by the inviscid contour-dynamics simulation of Dritschel
\& Waugh (1992), we seek to identify Lagrangian vortex cores in the
interaction of two vortices of unequal strength (cf. our discussion
in Section 10.1). We solve the Navier--Stokes equation \eqref{eq:NS}
in two dimensions with viscosity $\nu=2\times10^{-6}$. As initial
velocity field, we let 
\begin{align*}
v(x,0)=v_{1}(x)+v_{2}(x),\qquad v_{i}(x)= & \Gamma_{i}\frac{1-\exp(-r_{i}^{2}/\delta_{i}^{2})}{2\pi r_{i}^{2}}\,R_{\pi/2}\,(x-x_{i}),\quad i=1,2,
\end{align*}
where $r_{i}^{2}=|x-x_{i}|^{2}$ , and the matrix $R_{\pi/2}$ refers
counter-clockwise rotation by $\pi/2$. Each $v_{i}$ defines a vortex
centered at a point $x_{i}$ with a Gaussian profile. The parameters
$\Gamma_{i}$ and $\delta_{i}$ control the strength and width of
each vortex, respectively. In our simulation. we chose 
\begin{align*}
\Gamma_{1}=0.01,\quad\delta_{1}=0.05,\quad x_{1}=(\pi,\pi+0.2),\\
\Gamma_{2}=0.12,\quad\delta_{2}=0.20,\quad x_{2}=(\pi,\pi-0.2).
\end{align*}
The vortex described initially by $v_{1}$ is, therefore, weaker and
smaller, and hence is expected to deform significantly under the strain
field created by the second vortex initially described by $v_{2}$.
The computational domain is $x\in[0,2\pi]\times[0,2\pi]$, discretized
into $512\times512$ grid points. During the simulation, the vortices
stay far enough from the boundary so that the boundary effects on
their evolution are negligible. A relevant Reynolds number can be
computed as $Re=\Gamma_{2}/\nu=6\times10^{4}$ (cf. Kevlahan \& Farge
1997). For this Reynolds number, the flow still remains fairly close
to its inviscid limit, with the standard deviation of the vorticity
from its initial value staying below $0.5\%$ along fluid trajectories.

\begin{figure}
\centering
\subfloat[]{\includegraphics[width=0.4\textwidth]{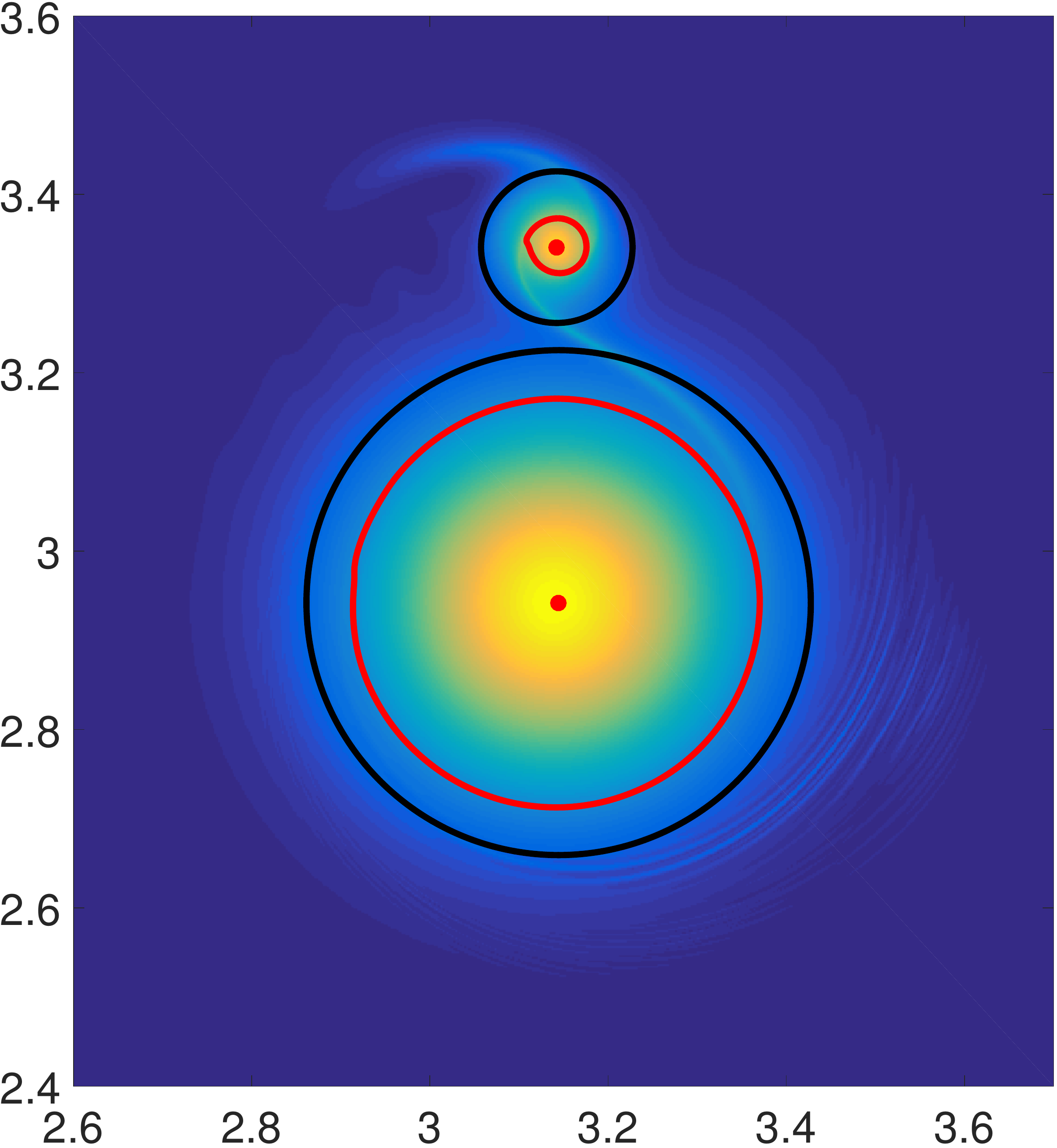}}\qquad
\subfloat[]{\includegraphics[width=0.4\textwidth]{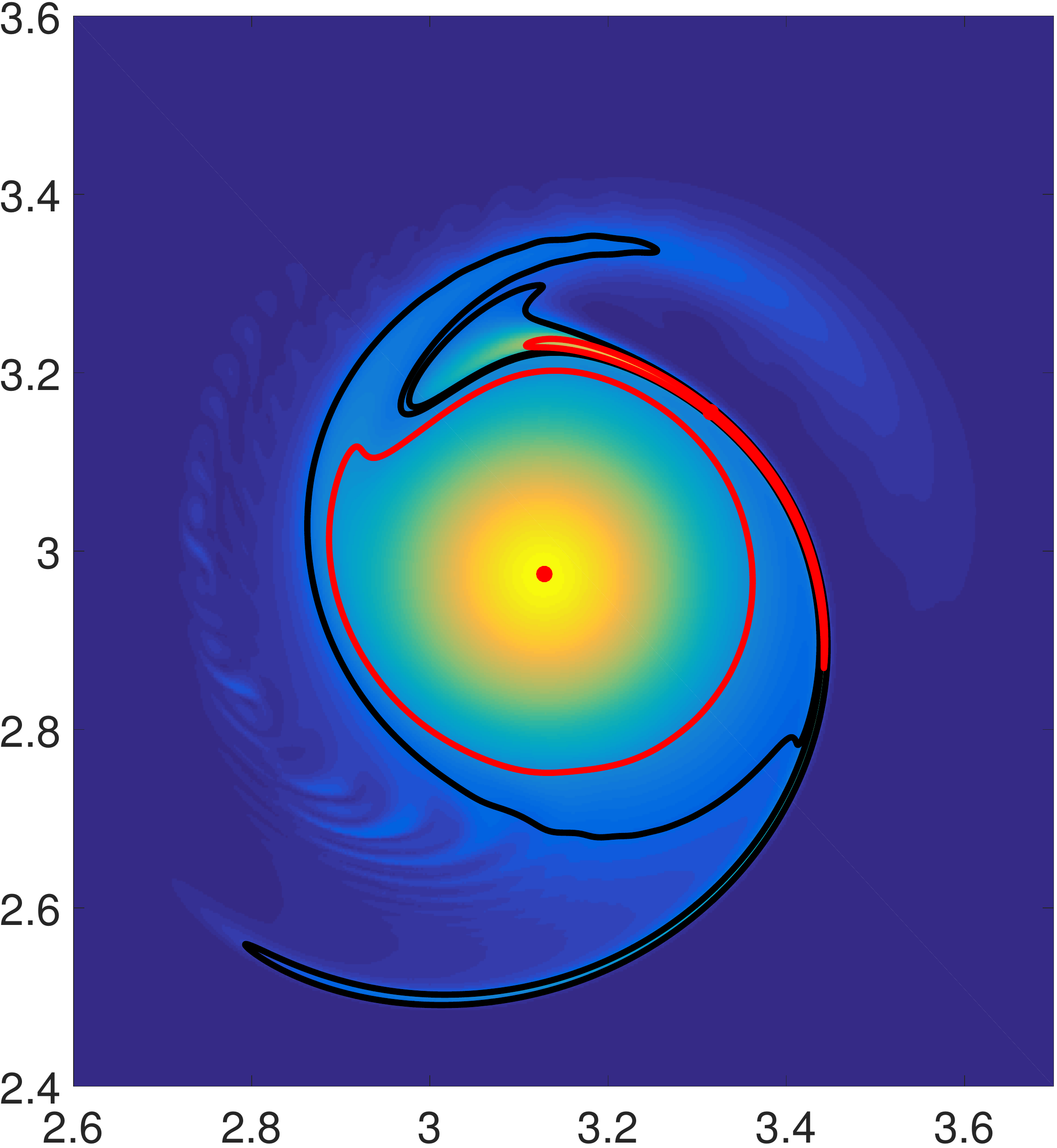}}
\caption{(a) Rotationally coherent Lagrangian vortex boundaries (red) at
time $t_{0}=0$ in the vortex interaction example. Also shown are
two circles (black) approximating the Eulerian vortex boundaries inferred
at time $t_{0}$ from the vorticity distribution. The $LAVD_{0}^{30}(x_{0})$
is shown in the background for reference. (b) advected position
of the Lagrangian vortex boundaries and the two closed material curves
at time $t=30.$ The advected LAVD field $LAVD_{0}^{30}(x(30;0,x_{0}))$
is shown in the background. (See the on-line supplemental movie M2
for the complete advection sequence of the vortex boundaries).}
\label{fig:vortex_interaction} 
\end{figure}

We have run the computation from the non-dimensional time $t_{0}=0$
up to $t=30.$ Over this time scale, we could preserve the smoothness
of the velocity field without additional numerical effort. We applied
Algorithm 1 with the tight convexity deficiency $d_{max}=10^{-3}$
and the arclength filter $l_{min}=0.1$. We show the rotationally
coherent vortex boundaries obtained in this fashion in Fig. \ref{fig:vortex_interaction},
along with two other closed material lines initiated around the vortex
cores. The latter two circles approximate closely the Eulerian vorticity
boundaries inferred from the instantaneous vorticity at time $t_{0}=0$.
They, therefore, mimic the the role of the initially circular vorticity
jump contours advected in the inviscid simulation of Dritschel \&
Waugh (1992).

As seen in Fig. \ref{fig:vortex_interaction}, the stronger Eulerian
vortex creates a sizable rotationally coherent Lagrangian vortex in
the center. This vortex shows only tangential filamentation, as confirmed
by its advected position at $t=30$. The weaker Eulerian vortex has
a substantially smaller coherent Lagrangian footprint that orbits
around the larger vortex.

This weaker satellite vortex gradually reaches a maximal elongated
perimeter at about $t=24$, then preserves its arclength in an approximate
rigid-body rotation for the remaining $20\%$ of the simulation time.
No transverse filamentation occurs in this case either: the satellite
vortex remains rotationally coherent in the sense of our Definition
1.

Therefore, as Figure \ref{fig:vortex_interaction} illustrates, two
unequal viscous vortices may interact strongly and still preserve
their rotationally coherent material cores during a finite time interval
of their interaction. Such coherent cores remain hidden in contour
dynamics studies focused on the advection of Eulerian vortex boundaries
inferred from large initial vorticity gradients. Indeed, the two black
material curves of Figure \ref{fig:vortex_interaction}, mimicking
the role of vortex-bounding inviscid vorticity contours, develop substantial
transverse filamentation during the same time interval.

\subsection{Two-dimensional Agulhas eddies in satellite altimetry\label{sub:Two-dimensional-geostrophic-flow}}

Here we illustrate the detection of rotationally coherent eddy boundaries
in velocity data derived from satellite-observed sea-surface heights
under the geostrophic approximation. In this approximation, the satellite-measured
sea-surface height $\eta(\varphi,\theta,t)$ serves as a non-canonical
Hamiltonian for surface velocities in $(\varphi,\theta)$ longitude-latitude
$(\varphi,\theta)$ coordinate system.

The evolution of fluid particles satisfies 
\begin{eqnarray}
\dot{\varphi}(\varphi,\theta,t) & = & -\frac{g}{R^{2}f(\theta)\cos\theta}\ \partial_{\theta}\eta(\varphi,\theta,t),\\
\dot{\theta}(\varphi,\theta,t) & = & \frac{g}{R^{2}f(\theta)\cos\theta}\ \partial_{\varphi}\eta(\varphi,\theta,t),
\end{eqnarray}
where $g$ is the constant of gravity, $R$ is the mean radius of
the Earth, and $f(\theta)\equiv2\Omega\sin\theta$ is the Coriolis
parameter, with $\Omega$ denoting the Earth's mean angular velocity.
The publicly available AVISO sea-surface height data base for $\eta(\varphi,\theta,t)$
is given at a spatial resolution of $1/4^{\circ}$ and a temporal
resolution of $7$ days.

We select the computational domain in the longitudinal range $[-4^{\circ},9^{\circ}]$
and the latitudinal range $[-35^{\circ},-28{}^{\circ}]$, which falls
inside the region of the Agulhas leakage in the Southern Ocean. A
Lagrangian analysis of coherent mesoscale eddies is particularly important
in this context, as the amount of warm and salty water carried from
the Indian Ocean to Atlantic Ocean has relevance for global circulation
and climate (Beal et al. 2011).

Recent two-dimensional Lagrangian studies of the Agulhas leakage used
geodesic LCS theory to locate perfectly non-filamenting (black-hole
type) material eddies (Haller \& Beron--Vera 2013, Beron--Vera et
al. 2013). Here we use a relaxed notion of coherence, which allows
for tangential filamentation, but not for global break-away of material
from the eddy. The computational cost in the present method is substantially
lower: the calculation of the deformation gradient and the search
for limit cycles bounding the black-hole eddies are absent.

We consider the AVISO data set ranging from the initial time $t_{0}=\mathrm{November}\,11,\,2006$
to the final time $t_{1}=t_{0}+90\,\mathrm{days}$. We select an initial
grid of particles with stepsize $\Delta x_{0}=1/50^{\circ}$. As an
additional filter for eliminating LAVD and IVD maxima due to resolution
coarseness, we ignore from the start LAVD and IVD maxima that are
closer than the submesoscale distance $0.2^{\circ}$ (about 20 km).
As a complexity deficiency bound, we select $d_{\max}=10^{-3}$. As
an arclength threshold, we fix $l_{\min}=2\pi r$, with $r=20$\,km
selected again as a lower bound on mesoscale structures reliably resolved
by altimetry.

\begin{figure}
\centering
\includegraphics[width=0.6\textwidth]{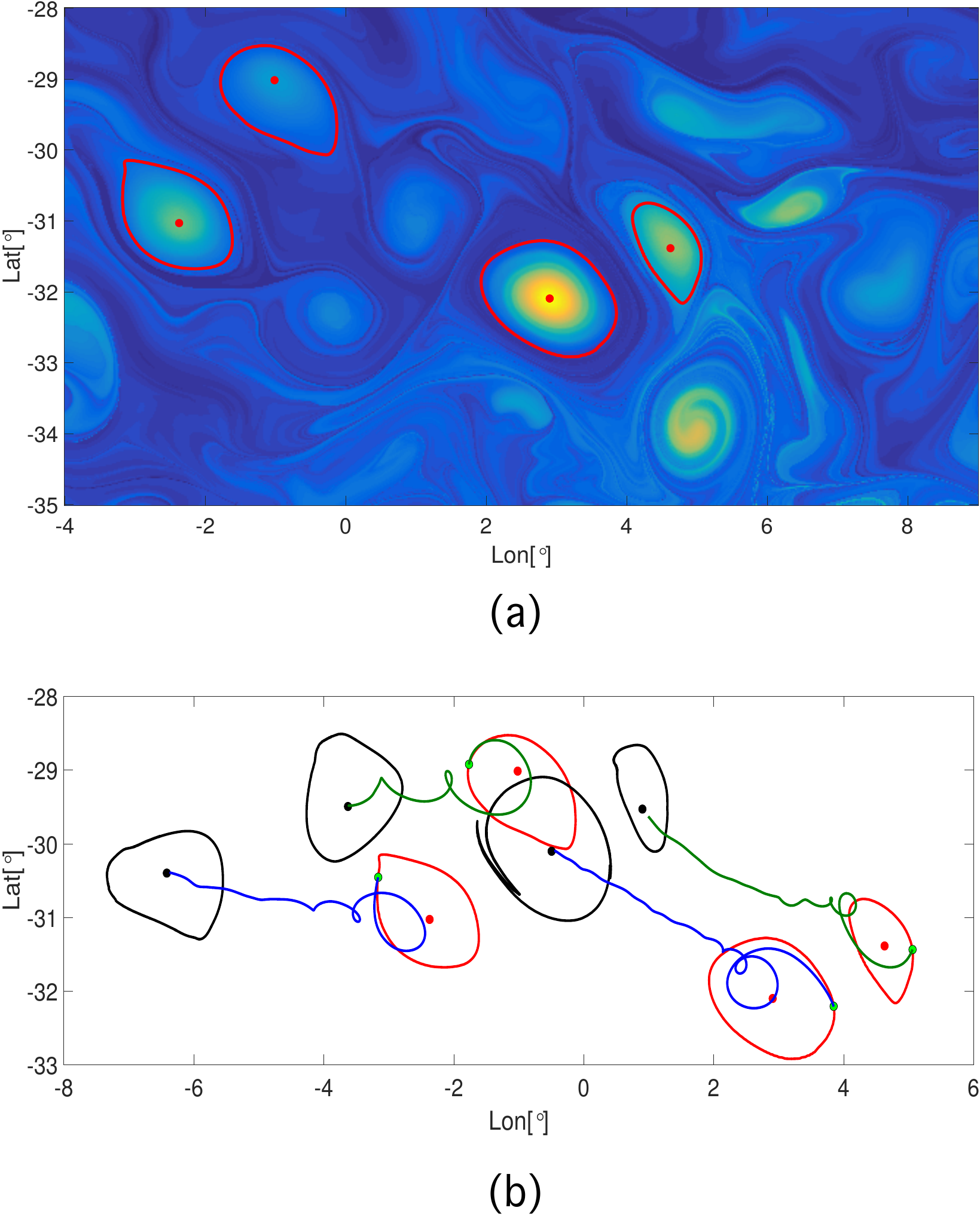}
\caption{(a) Rotationally coherent Lagrangian vortices at time $t_{0}=\mathrm{November}\,11,\,2006$,
identified from Algorithm 1 using the contours of $\mathrm{LAVD}_{t_{0}}^{t_{0}+T}(x_{0})$
with $T=90\,\mathrm{days}$. Shown in the background is the contour
plot of $\mathrm{LAVD}_{t_{0}}^{t_{0}+T}(x_{0})$ for reference. (b)
Initial (red) and final (black) positions of the Lagrangian vortex
boundaries at time $t_{0}+T$, along with representative inertial
particle trajectories. Heavy particles (blue) converge to the centers
of anti-cyclonic (clockwise) eddies. Light particles (green) converge
to the centers of cyclonic (clockwise) eddies. Here $\delta_{heavy}=0.99$,
$\delta_{light}=1.01$, $r_{0}=1$\,m. (See the on-line supplemental
movie M3 for the complete advection sequence of the vortex boundaries
and the inertial particles).}
\label{fig:geostrophy+inertial} 
\end{figure}

Beyond executing Algorithm 1 to extract rotationally coherent in this
setting, we also use this example to illustrate the predictions obtained
from Theorem 2 for the attractor role of rotationally coherent Lagrangian
vortex centers. Figure \ref{fig:geostrophy+inertial}a shows the rotationally
coherent eddies, while Figure \ref{fig:geostrophy+inertial}b confirms
that their boundaries only develop tangential filamentation under
Lagrangian advection, as expected. This second plot also confirms
that LAVD-based vortex centers are precisely the observed attractors
for light particles released in cyclonic eddies, and for heavy particles
released in anticyclonic eddies.

\begin{figure}
\centering
\includegraphics[width=1\textwidth]{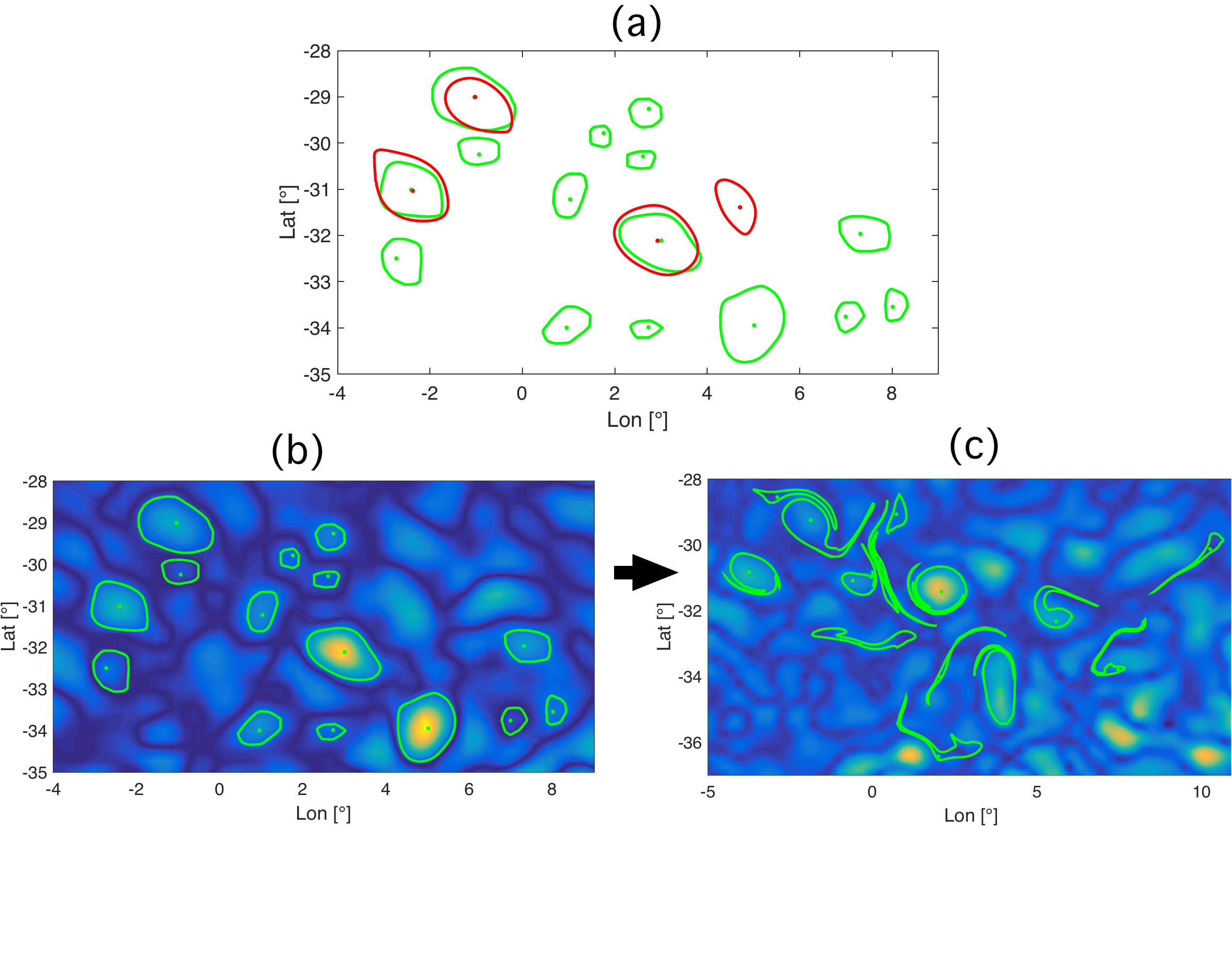}
\caption{(a) Lagrangian (red) and Eulerian (green) rotationally coherent vortices
at time $t_{0}$ in the satellite altimetry example. (b) Rotationally
coherent Eulerian vortex boundaries and centers at time $t_{0}$,
with the $\mathrm{IVD}(x,t_{0})$ field shown in the background. (c)
The same objects advected passively to the final time $t_{1}=t_{0}+T$,
with $\mathrm{IVD}(x,t_{1})$ shown in the background. }
\label{fig:altimetry_IVD} 
\end{figure}

We show a comparison of the initial positions of rotationally coherent
Lagrangian and Eulerian vortices in Fig. \ref{fig:altimetry_IVD}a.
Three of the Eulerian eddies (green) are close to Lagrangian eddies
(red), and will accordingly show some coherence under advection by
the end of the observational period in Fig. \ref{fig:altimetry_IVD}c.
The remaining Eulerian eddies show major filamentation and disintegrate
under advection. We note the large number of false positives for coherent
eddies based on the instantaneous Eulerian prediction at the initial
time, even though this prediction is frame-invariant.

In Appendix D, we also use this example to illustrate differences
between LAVD-based vortex detection and two other objective Lagrangian
tools for two-dimensional flows: the geodesic LCS approach of Haller
\& Beron--Vera (2013) and the ellipticity-time diagnostic of Haller
(2001).

We find that the geodesic LCS approach generally identifies similar
material vortex regions as LAVD. By construction, however, geodesic
vortex boundaries enclose perfectly non-filamenting vortex cores,
and hence miss the rotationally coherent (but tangentially filamenting)
outer annuli of Lagrangian vortices. The identification of the perfectly
coherent, black-hole-type core by the geodesic LCS approach also comes
with a higher computational cost and does not extend to three dimensions
(cf. Appendix D for details).

We also find that the ellipticity-time diagnostic highlights the general
vicinity of the coherent material vortex regions, but offers no well-defined
procedure for defining vortex boundaries. Focused on individual trajectory
stability rather than global coherence, the ellipticity time has equally
high values in some regions that do not remain coherent as a whole.
Conversely, the ellipticity time indicates predominantly hyperbolic
(i.e., non-vortical) behavior in the outer, tangentially filamenting
parts of rotationally coherent vortices.

In summary, despite their higher computational costs, neither globally
uniform-stretching material lines nor elliptic fluid trajectories
are able to provide the large, coherent material vortex boundaries
obtained from the LAVD.

\section{Three-dimensional examples }

\subsection{Strong Beltrami flows}

Strong Beltrami flows are steady flows whose vorticity field is a
constant scalar multiple of the velocity field (Majda \& Bertozzi
2002), i.e. 
\[
\omega(x)=\lambda v(x).
\]
In this case, formula \eqref{eq:LAVD} gives 
\begin{equation}
\left.\mathrm{LAVD}\right._{t_{0}}^{t}(x_{0})=\left|\lambda\right|\intop_{t_{0}}^{t}\left|v(x(s))-\bar{v}\right|ds,\qquad\bar{v}:=\frac{1}{\mathrm{vol}\,(U)}\int_{U}v(x)\,dV.
\label{eq:LAV-Beltrami}
\end{equation}

Thus, for any strong Beltrami flow with $\lambda\neq0$, \emph{a rotationally
coherent Lagrangian vortex boundary is a locally outermost, closed
and convex level surface of} $\intop_{t_{0}}^{t}\left|v(x(s))-\bar{v}\right|ds,$
the trajectory-averaged deviation of the velocity field from its spatial
mean. For general 3D steady flows, level sets of asymptotically Lagrangian-averaged
observables have been noted to approximate ergodic components (Budi\v{s}i\'{c}
\& Mezi\'{c} 2012). The new result here is that tubular levels sets
of the finite-time average of the normed velocity deviation in strong
Beltrami flows specifically define rotationally coherent Lagrangian
structures (surfaces of constant net bulk rotation) in an objective
fashion.

By formula \eqref{eq:LAV-Beltrami}, rotationally coherent Eulerian
vortices in strong Beltrami flows are composed of convex tubular level
surfaces of the normed velocity deviation $\mathrm{IVD}(x)=$$\left|v(x)-\bar{v}\right|$.

As an example, we consider the ABC flow whose velocity field is given
by 
\[
v(x)=\left(\begin{array}{c}
A\sin x_{3}+C\cos x_{2}\\
B\sin x_{1}+A\cos x_{3}\\
C\sin x_{2}+B\cos x_{1}
\end{array}\right)
\]
over the triply periodic domain $U=[0,2\pi]^{3}$. A direct calculation
gives $\bar{v}(x)\equiv0$ and $\lambda=1$. We select the parameter
configuration $A=1,$ $B=\sqrt{2/3}$, and $C=\sqrt{1/3}$. For our
Lagrangian advection, we select an initial grid of $200^{3}$ evenly
distributed particles, which we advect over the interval $[t_{0},t_{1}]=[0,50]$.

\begin{figure}
\centering
\includegraphics[width=1\textwidth]{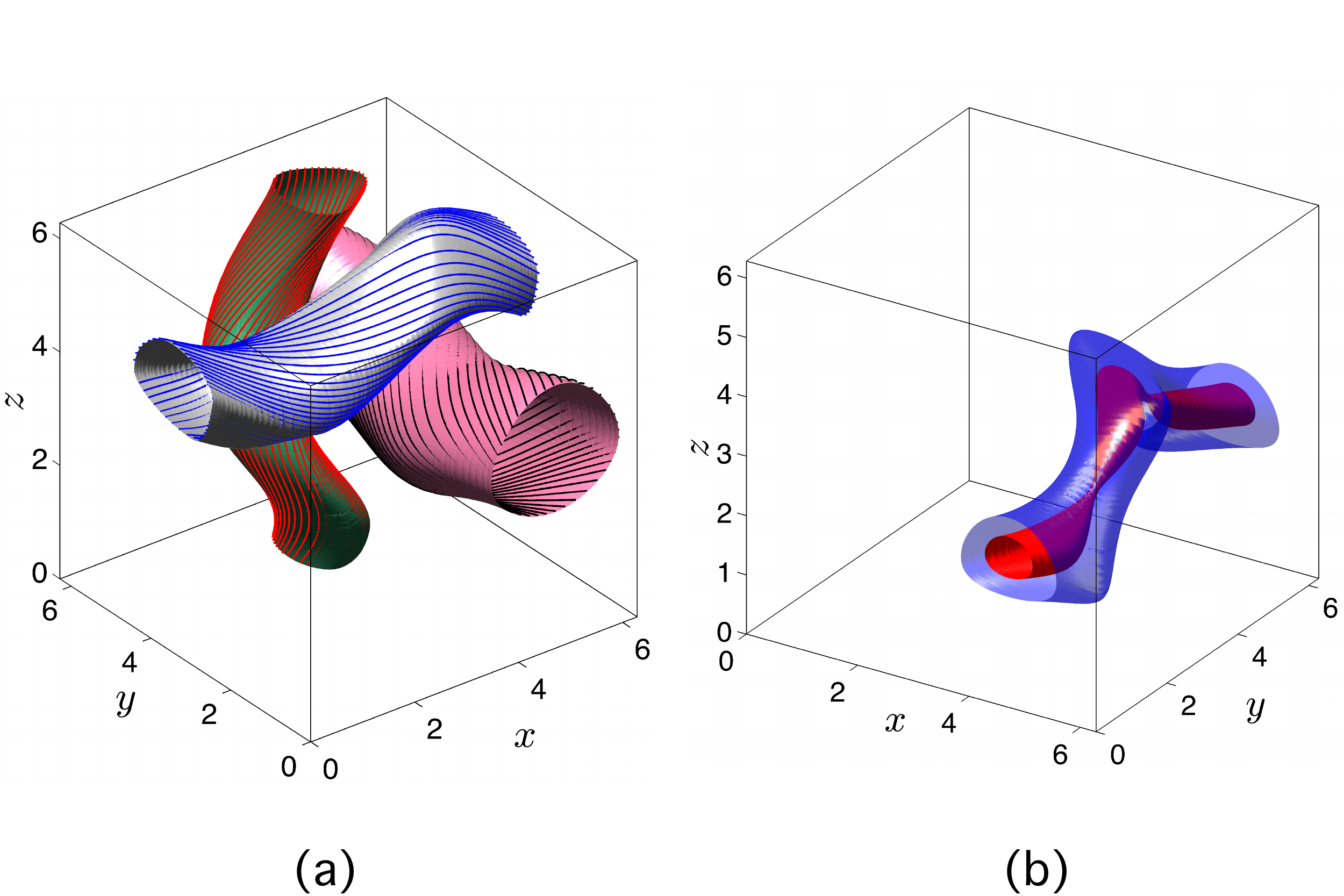}
\caption{(a) Three representative rotationally coherent Lagrangian vortex boundaries
in the ABC flow, obtained as outermost, convex tubular level sets
of $\left.\mathrm{LAVD}\right._{0}^{50}(x_{0})$. Black curves indicate
trajectories launched on these boundaries. (b) Two representative
$\mathrm{IVD}(x_{0})$ level surfaces (with blue denoting the rotationally
coherent Eulerian vortex boundary) in the single Eulerian coherent
vortex obtained from the IVD field for the ABC flow for the parameter
values considered here.}
\label{fig:ABCflow} 
\end{figure}

Figure \ref{fig:ABCflow}a shows three representative Lagrangian vortex
boundaries obtained as outermost, convex tubular level surfaces of
$\left.\mathrm{LAVD}\right._{0}^{50}(x_{0})$. By the simplicity of
this steady flow, relaxing convexity was not necessary (the convexity
deficiency is zero). Also shown in Figure \ref{fig:ABCflow} are trajectory-segments
launched along the Lagrangian vortex boundaries, illustrating their
Lagrangian invariance. Figure \ref{fig:ABCflow}b shows the single
IVD-based Eulerian vortex obtained for the ABC flow. While this Eulerian
structure is near one of the Lagrangian vortices, the shape of the
IVD-based vortex is not representative of the true Lagrangian vortex
in this example. Furthermore, no IVD-based Eulerian vortices arise
near the remaining Lagrangian vortices. This might seem puzzling first,
but one must remember the spatial anisotropy of the ABC flow for unequal
$A,B$ and $C$ parameter values.

\subsection{Three-dimensional Agulhas eddies in a data-assimilating circulation
model}

\label{Ex:SOSE} Here we apply LAVD-based vortex extraction to a three-dimensional
unsteady velocity field set obtained from the Southern Ocean State
Estimation (SOSE) model (Mazloff, Heimbach \& Wunsch 2012). The domain
of the data set is again in the area of the Agulhas leakage in the
Southern Ocean, representing a three-dimensional extension of the
Agulhas eddy extraction study of Section \ref{sub:Two-dimensional-geostrophic-flow}.

Our Lagrangian study covers a period of $T=120\,\mathrm{days}$, ranging
from $t_{0}=$ 15 May 2006 to $t=$ 12 September 2006. The selected
computational domain in the South Atlantic ocean is bounded by longitudes
$[11^{\circ}\text{E},16^{\circ}\text{E}]$, latitudes $[37^{\circ}\text{S},33^{\circ}\text{S}]$,
and depths $[7,2000]$ meters. We compute the LAVD and IVD fields
over a uniform grid of $350\times350\times350$ points, and identify
a rotationally coherent Lagrangian and Eulerian vortices using Algorithm
\ref{alg:algorithm2}. As in our two-dimensional turbulence example,
we integrate the vorticity deviation norm separately (as opposed to
solving the combined ODE \eqref{eq:combined_ODE}), using 1000 vorticity
values along each trajectory, equally spaced in time. For the plane
family $\mathcal{P}_{i}$ featured in Algorithm \ref{alg:algorithm2},
we consider horizontal planes along nodes of the initial grid, starting
from 28 meters below the sea surface level to eliminate noise due
to boundary effects at the surface. In these planes we use the arclength
threshold $l_{\min}=0.1$ and the maximal convexity deficiency $d_{\max}=10^{-4}.$
The spatial mean vorticity $\bar{\omega}(t)$, as the practically
observed mean vorticity for a large enough fluid mass in the ocean,
is taken to be zero.

\begin{figure}[h!]
\centering 
\subfloat[]{\includegraphics[width=0.8\textwidth]{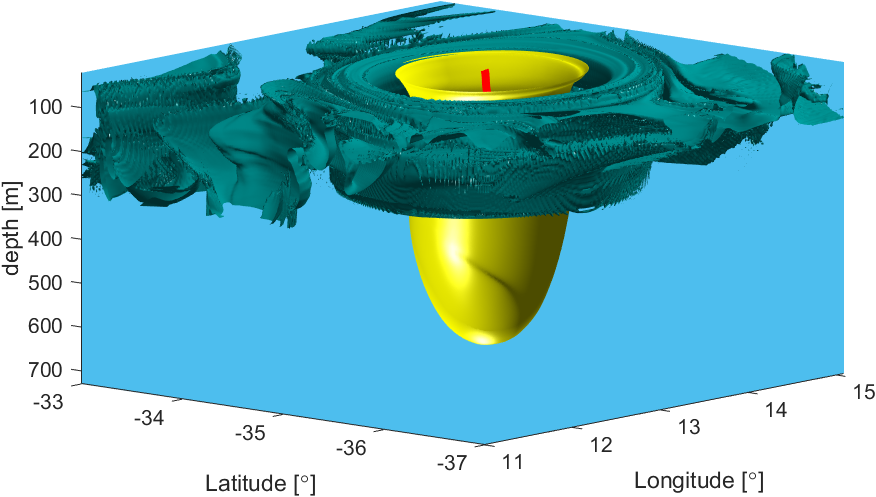}}\\
\subfloat[]{\includegraphics[width=0.4\textwidth]{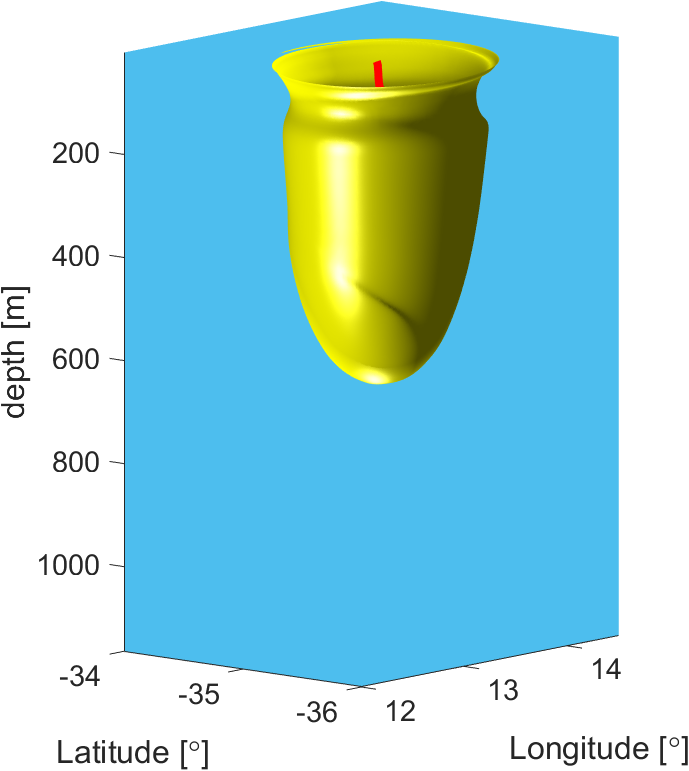}}
\subfloat[]{\includegraphics[width=0.4\textwidth]{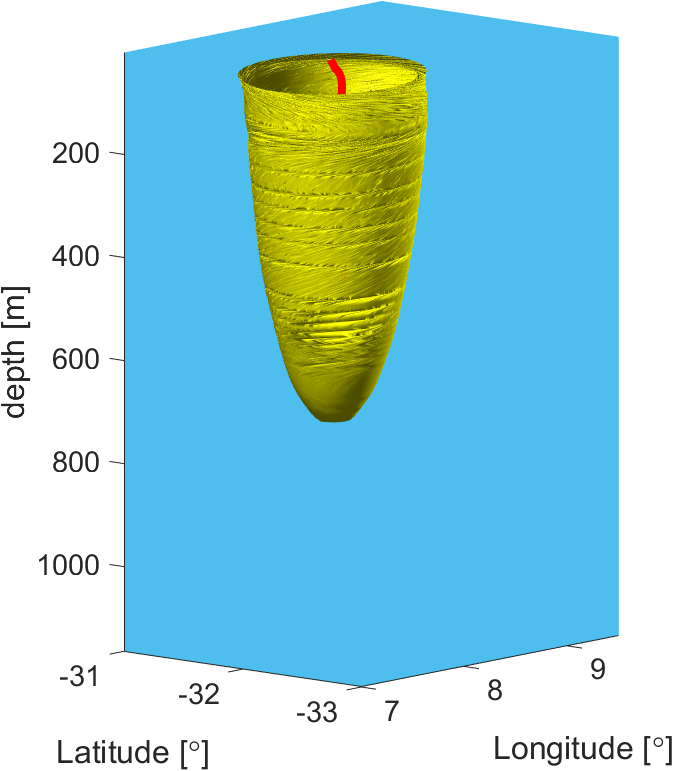}}
\caption{(a) Representative level surfaces of $\mathrm{LAVD}_{t_{0}}^{t_{0}+T}(x_{0})$
with $t_{0}=\mathrm{May}\,15,\,2006$ and $T=120$ days. ~The yellow
surface is extracted using Algorithm 2, marking the vortex boundary
for a mesoscale, rotationally coherent Lagrangian eddy, extending
from $28$ meters down to $646$ meters in depth. The green surface
is a nearby level surface of $\mathrm{LAVD}_{t_{0}}^{t_{0}+T}(x_{0})$
outside the Lagrangian vortex region. The red curve marks the coherent
vortex center, as defined in Definition 1. (b) Full view of the Lagrangian
eddy boundary and its center at the initial time $t_{0}$ (c) The
advected eddy boundary and vortex center $120$ days later, extending
from 49 meters down to 726 meters below the surface (See the on-line
supplemental movie M4 for the complete advection sequence of the eddy
boundary).}
\label{fig:R1SOSE3D} 
\end{figure}

Figure \ref{fig:R1SOSE3D}a shows the initial position of a rotationally
coherent Lagrangian eddy boundary (yellow) and its center (red), extracted
as a level sets of $\mathrm{LAVD}_{t_{0}}^{t_{1}}(x_{0})$ by Algorithm
2. Also shown is a nearby LAVD level surface outside the eddy boundary,
illustrating the complexity of the near-surface mixing region enclosing
the eddy. Figure \ref{fig:R1SOSE3D}b gives a full view of the Lagrangian
eddy, whereas Figure \ref{fig:R1SOSE3D}c shows the materially advected
position of the eddy at the final time, $120$ days later. As expected,
there is mild tangential filamentation in the material eddy boundary,
but strictly no break-away from the rotating water mass. Given the
complexity of material mixing in the surrounding waters, this high
degree of material coherence illustrates well the accuracy of LAVD-based
vortex extraction.

\begin{figure}
\centering
\subfloat[]{\includegraphics[width=0.4\textwidth]{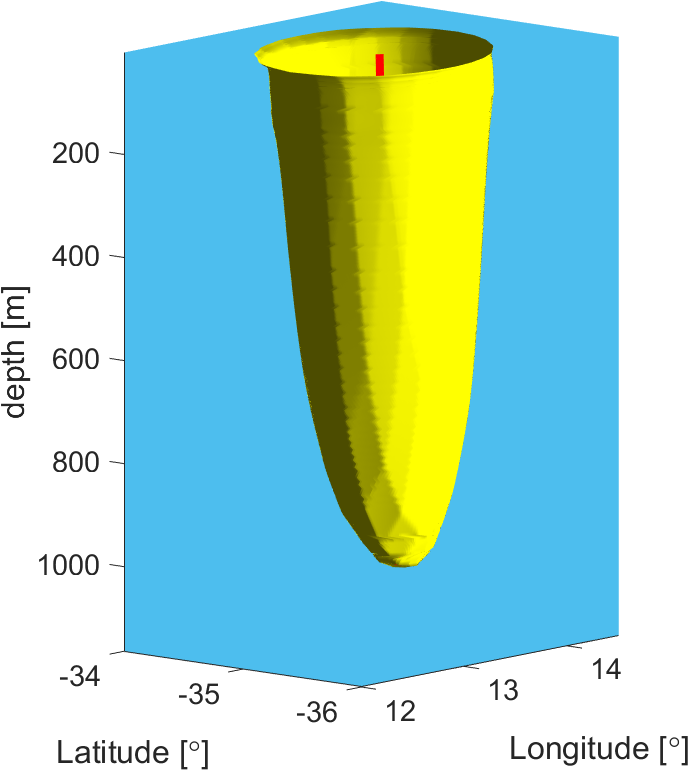}}
\subfloat[]{\includegraphics[width=0.4\textwidth]{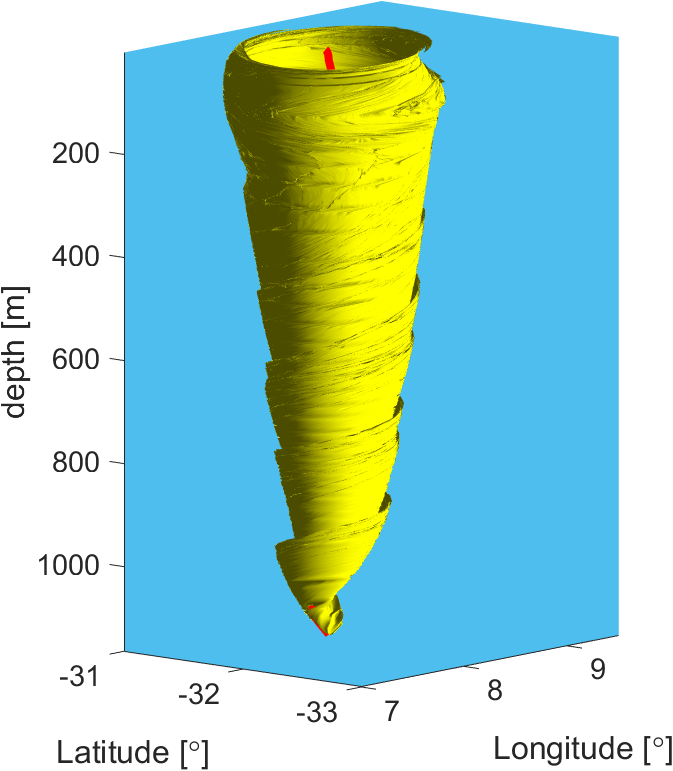}}
\caption{(a) The rotationally coherent Eulerian counterpart of the Lagrangian
eddy shown in Figure \ref{fig:R1SOSE3D}a, ranging from 7 $m$ down
to 1,007 $m$ below the see surface. (b) The materially advected position
of the Eulerian eddy 120 days later, ranging from 15 $m$ to 1,163
$m$ below the sea surface (See the on-line supplemental movie M5
for the complete advection sequence of the eddy boundary).}
\label{fig:R1SOSE3D_IVD} 
\end{figure}

The corresponding rotationally coherent Eulerian eddy extracted at
the same initial location, then materially advected for 120 days,
is shown in Fig. \ref{fig:R1SOSE3D_IVD}. In the Eulerian computation,
noise in the level surface computation is more moderate than in the
Lagrangian case. As a result, the plane family $\mathcal{P}_{i}$
featured in Algorithm \ref{alg:algorithm2} can be selected to start
from as high as 7 meters below the surface.

This Eulerian eddy has about the same diameter near the sea surface
as its Lagrangian counterpart, but maintains this diameter and reaches
to substantially larger depths. Its vertical size is further increased
under advection, with the bottom forming a sharp tip. Overall, the
advected surface shows high ribbing and filamentation. Large-scale
material break-away is absent in this example, but this cannot be
guaranteed a priori for an Eulerian eddy (see our two-dimensional
computations in Section \ref{sub:Two-dimensional-geostrophic-flow}).
A comparison of Figures \ref{fig:R1SOSE3D} and \ref{fig:R1SOSE3D_IVD}
suggests that the Lagrangian eddy forms a smooth, coherent center
region that transports water without observable filamentation inside
the vortical Eulerian feature.

\begin{figure}
\centering 
\includegraphics[width=1\textwidth]{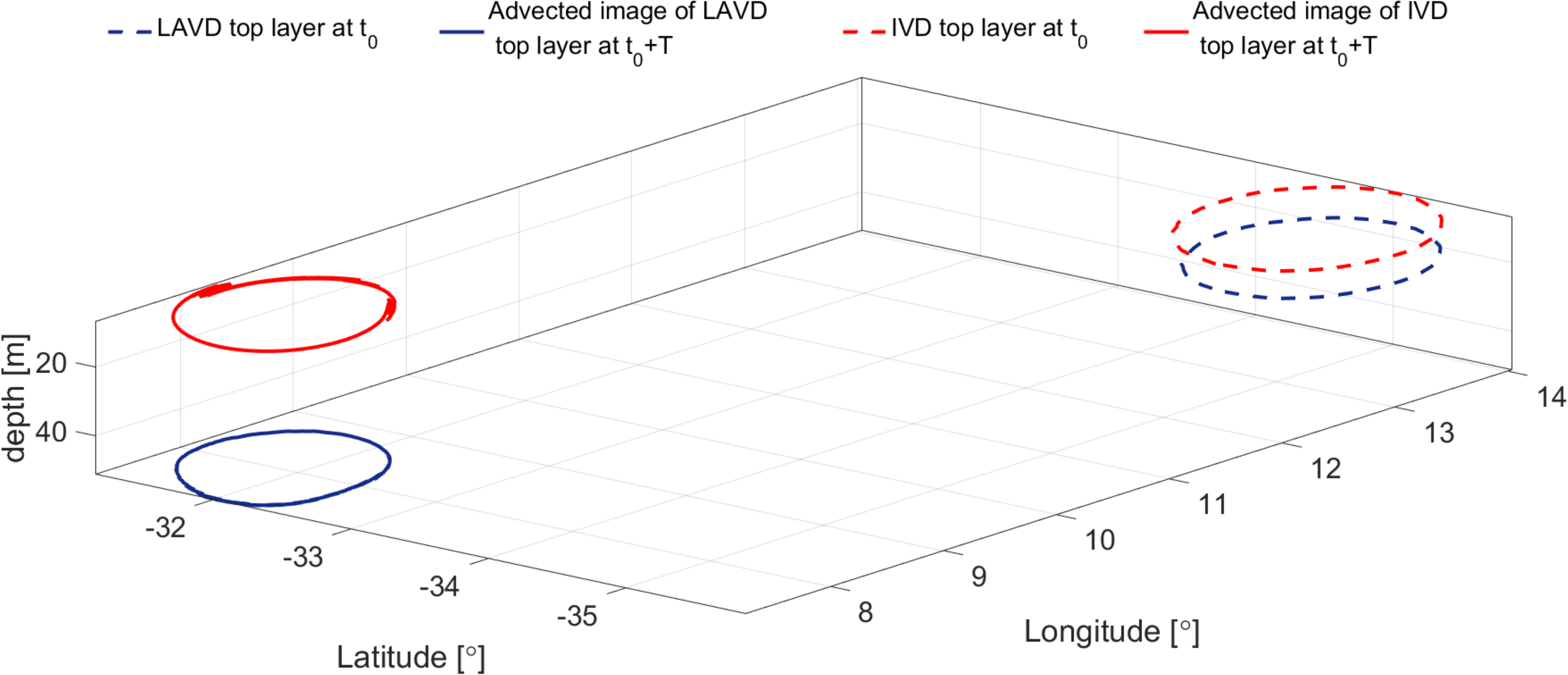}
\caption{Initial top slices (dashed line, at time $t_{0}=\mathrm{May}\,15,\,2006$)
and their advected positions (solid line, $T=120\,days$ later) of
the LAVD- and IVD-based rotationally coherent vortex boundaries.}
\label{fig:R1SOSE3D_LAVD_top} 
\end{figure}

With a larger number of dedicated control points,
the computation of LAVD level surfaces can be extended closer to the
ocean surface. Figure \ref{fig:R1SOSE3D_LAVD_top} shows a higher-resolution
computation of the initial and advected positions of the top slices
of IVD-based and LAVD-based rotationally coherent vortex boundary
surfaces. In this computation, the initial top slice of the LAVD-based
vortex boundary is located only $15$ $m$ below the ocean surface,
as opposed to the $28$ $m$ distance used in Fig. \ref{fig:R1SOSE3D}c.
Even this Lagrangian boundary slice remains more coherent than the
Eulerian one, but the Eulerian slice still performs well under material
advection for this eddy and for this advection time.

\section{Conclusions}

We have given an objective (fully observer-independent) definition
of a Lagrangian vortex as a set of material tubes in which fluid elements
complete the same intrinsic dynamic rotation. This material rotation
angle is obtained from the exact, dynamically consistent decomposition
\eqref{eq:full_decomp} of the deformation gradient. Remarkably, the
intrinsic material rotation angle is expressible as the trajectory
integral of the normed deviation of the vorticity from its spatial
mean (LAVD). The intrinsic material rotation is, therefore, directly
observable as the rotation of vorticity-meters in fluid experiments
(Shapiro 1961), once the mean rotation reported by these devices is
subtracted from the measurements.

Locating a rotationally coherent material vortex does not require
advection of high-density material grids, a generally taxing numerical
procedure in Lagrangian coherence calculations. By construction, LAVD-based
material vortices may show material filamentation, but even filamented
material elements will rotate together with the vortex without global
breakaway. The vortex interior, therefore, shows no advective mixing
with its environment, as we have illustrated on several several examples
in two- and three-dimensional unsteady flows.

Our approach also enables the extraction of vortex centers as singular
level sets (local maxima) of the LAVD. We have proved that in two-dimensional
geostrophic flows, these vortex centers coincide precisely with attractors
of light particles in cyclonic eddies and for heavy particles in anticyclonic
eddies. Indeed, we have found that numerically simulated light and
heavy inertial particles show quick convergence to the appropriate
moving vortex cores in a satellite-inferred, geostrophic ocean velocity
field. On the same example, we have illustrated the advantages of
LAVD-based coherent vortex detection compared to other objective Lagrangian
vortex detection tools.

The deviation of potential vorticity from its spatial mean is tempting
to consider for a similar analysis, but such an approach would remain
heuristic, driven purely by analogy. This is because potential vorticity
is not objective and cannot be rigorously connected to intrinsic material
rotation generated by the deformation gradient.

Motivated by our results on LAVD, one might proceed by analogy and
probe plots of Lagrangian averages of arbitrary scalar fields for
vortical features. As long as these scalar fields are non-degenerate
and favorably initialized, material vortices should indeed have a
footprint in the resulting plots due to the coherence of trajectories
in the vortex interiors. For instance, the trajectory-averaged Okubo--Weiss
parameter (Dresselhaus \& Tabor 1989, P�rez--Mu�uzuri \& Huhn 2013)
and trajectory-averaged helicity (P�rez--Mu�uzuri \& Huhn 2013) may
also have lows and highs, respectively, near material vortices. Such
extrema are, however, heuristic diagnostics, each signaling a different
domain for a vortex, and each showing extrema in other flow regions
as well. Their features are a consequence, rather than a well understood
root cause, of material coherence (cf. Beron--Vera 2015).

In contrast, we arrive here at LAVD-based vortices by solving an objectively
posed coherence problem for material surfaces of equal bulk rotation.
The LAVD then arises from an exact, dynamically consistent decomposition
of the deformation gradient into purely rotational and purely straining
deformation gradients. The vortex boundaries and centers so obtained
are sharply defined, do not develop global material filamentation,
and remain invariant with respect to all possible Euclidean observer
changes.

We have also formulated an objective Eulerian definition of a rotationally
coherent vortex: a domain filled with tubular surfaces of constant
intrinsic material rotation rate. These surfaces coincide with outward-decreasing
tubular level sets of the instantaneous vorticity deviation (IVD).
In some cases, we have found rotationally coherent Eulerian vortices
and their centers to be surprisingly close to their Lagrangian counterparts.
While such closeness does not hold in general, IVD-based vortices
do provide a systematic and frame-invariant way to track coherent
velocity features that are infinitesimally consistent in time with
coherent material vortices. This makes these Eulerian vortices and
vortex centers appropriate tools for a fully frame-invariant, automated
vortex census in turbulent flow data.

\textbf{Acknowledgements}

We acknowledge helpful discussions with Matt Mazloff, and exploratory
work by Dan Blazevski on the SOSE data set. We are grateful to one
of the anonymous reviewers of this paper for suggesting the example
considered in Section 10.4 and the comparison carried out in Appendix
D. The altimeter products used in this work are produced by SSALTO/DUACS
and distributed by AVISO, with support from CNES (http://www.aviso.oceanobs.com/duacs).\\
 \\

\section{Appendix A: Material rotation from the classical Polar Decomposition\label{sec:PD}}

Finite strain theory (Truesdell \& Noll 1965) infers a unique bulk
material rotation for material volume elements between times $t_{0}$
and $t_{1}$ from the polar decomposition 
\begin{equation}
F_{t_{0}}^{t_{1}}=R_{t_{0}}^{t_{1}}U_{t_{0}}^{t_{1}},\qquad R_{t_{0}}^{t_{1}}=F_{t_{0}}^{t_{1}}\left[C_{t_{0}}^{t_{1}}\right]^{-1/2},\qquad U_{t_{0}}^{t_{1}}=\left[C_{t_{0}}^{t_{1}}\right]^{1/2},\label{eq:polardecomp}
\end{equation}
of the deformation gradient. Here the rotation tensor $R_{t_{0}}^{t_{1}}$
is proper orthogonal, interpreted as the solid-body rotation component
of the deformation. The right stretch tensor, $U_{t_{0}}^{t_{1}}$,
is symmetric and positive definite, obtained as the principal square
root of the Cauchy--Green strain tensor $C_{t_{0}}^{t_{1}}$. The
right stretch tensor is interpreted as the stretching preceding the
solid-body rotation represented by $R_{t_{0}}^{t_{1}}$. The tensors
$R_{t_{0}}^{t_{1}}$ and $U_{t_{0}}^{t_{1}}$ are not objective, but
the eigenvalues of $U_{t_{0}}^{t_{1}}$ are preserved under time-dependent
rotations and translations (Truesdell \& Noll 1965).

In a given frame of reference, $R_{t_{0}}^{t_{1}}(x_{0})$ represents
the unique rotation that gives the closest fit to the linear operator
$F_{t_{0}}^{t_{1}}(x_{0})$ in the Frobenius matrix norm (Gollub \&
Van Loan 1983). In two and three dimensions, the action of $R_{t_{0}}^{t_{1}}(x_{0})$
can be described by a single polar rotation angle (PRA). Well-defined
up to multiples of $2\pi$, PRA is the signed angle of rotation along
the axis of rotation associated with the tensor $R_{t_{0}}^{t_{1}}(x_{0})$

In recent work, we used tubular and singular level surfaces of the
PRA to visualize elliptic Lagrangian regions and their centers, respectively
(Farazmand \& Haller 2016). To our knowledge, this represents the
first systematic approach to identifying Lagrangian coherence based
on a synchrony in the net material rotation of infinitesimal volume
elements. The level sets of PRA are objective in two dimensions, and
have shown themselves to be accurate indicators of elliptic regions
in unsteady flows with general time dependence.

Nevertheless, several challenges remain that the classic polar decomposition,
and hence also the PRA, are unable to address: 
\begin{enumerate}
\item PRA level sets are not frame-invariant in three dimensions. Therefore,
PRA in three dimensions cannot be used to define elliptic LCSs, which
are material (and hence fundamentally frame-invariant) surfaces. 
\item The computation of the PRA is based on the invariants of the Cauchy--Green
strain tensor. This requires the accurate differentiation of the flow
map $\mathcal{F}_{t_{0}}^{t_{1}}(x_{0})$ with respect to initial
conditions. This is numerically costly for extended times in large
flow domains. 
\item No straightforward relationship exists between finite material rotation
represented by the PRA and physical quantities (most notably, vorticity)
used in Eulerian vortex identification. Indeed, only an involved link
exists between $R_{t_{0}}^{t_{1}}$ and the spin tensor $W$ through
the formula 
\begin{equation}
\dot{R}_{t_{0}}^{t}=\left(W-\frac{1}{2}R_{t_{0}}^{t}\left[\dot{U}_{t_{0}}^{t}\left(U_{t_{0}}^{t}\right)^{-1}-\left(U_{t_{0}}^{t}\right)^{-1}\dot{U}_{t_{0}}^{t}\right]\left(R_{t_{0}}^{t}\right)^{T}\right)R_{t_{0}}^{t},\label{eq:polar_rot_ODE}
\end{equation}
where dot refers to differentiation with respect to $t$ (Truesdell
\& Rajagopal 2009). 
\item Between two fixed times $t_{0}$ and $t_{1}$, the polar rotation
tensor $R_{t_{0}}^{t_{1}}$ represents the closest solid-body rotation
to $F_{t_{0}}^{t_{1}}$ in the Frobenius norm. For intermediate times
$t\in[t_{0},t_{1}],$the rotation family $R_{t_{0}}^{t}$ provides
no self-consistent solid-body rotation component for the evolving
deformation gradients $F_{t_{0}}^{t}$ . Indeed, one generally has
\begin{equation}
R_{t_{0}}^{t}\neq R_{s}^{t}R_{t_{0}}^{s},\qquad s,t\in[t_{0},t_{1}],\label{eq:no_product}
\end{equation}
which means that the rotation family $R_{t_{0}}^{t}$ does not satisfy
the basic superposition principle of subsequent rigid-body rotations.
Consequently, experimentally observed finite material rotation in
fluids (visualized by small rigid-body tracers (Shapiro 1961) will
differ from the PRA even in the simplest flows (see Haller 2016 for
further discussion). 
\end{enumerate}
Haller (2016) also discusses the rotational component yielded by $R_{t_{0}}^{t}$
for two simple fluid-mechanical examples: irrotational vortices and
parallel shear flows. In these examples, observed material rotation
signaled by small inertial tracers (vorticity-meters) differs fundamentally
from the rotation captured by $R_{t_{0}}^{t}$.

\section{Appendix B: Rotationally coherent vortices in flow over a moving
surface}

Here we show how the LAVD and IVD can be computed over moving surfaces,
such as the rotating Earth, without having to compute vorticity in
the curvilinear coordinates. Consider first a three-dimensional spatial
domain $U(t)\subset\mathbb{R}^{3}$ that possibly also translates
and rotates in time. We assume that $U(t)$ admits a globally orthogonal
parametrization

\begin{eqnarray}
f_{t}\colon\mathcal{U} & \subset & \mathbb{R}^{3}\rightarrow U(t),\label{eq:pardef}\\
\alpha & \mapsto & x,\nonumber 
\end{eqnarray}
where $\alpha$ is non-dimensionalized. If the velocity field at points
$x\in U(t)$ is denoted by $v(x,t),$ then 
\[
\dot{x}=v(f_{f}(\alpha),t)=Df_{t}(\alpha)\dot{\alpha}+\partial_{t}f(\alpha),
\]
which induces the corresponding velocity field 
\begin{equation}
\dot{\alpha}=\tilde{v}(\alpha,t):=\left[Df_{t}(\alpha)\right]^{-1}\left[v(f_{f}(\alpha),t)-\partial_{t}f(\alpha)\right]\label{eq:parflow}
\end{equation}
in the parameter space $\mathcal{U}$. The vorticity associated with
the flow \eqref{eq:parflow} in the parameter space is then given
by 
\begin{equation}
\tilde{\omega}(\alpha,t)=\nabla_{\alpha}\times\tilde{v},\label{eq:parvort}
\end{equation}
with $\nabla_{\alpha}$ denoting the gradient operation in the orthogonal
coordinates $\alpha.$ Theorem 1 is then applicable to the pull-back
flow \eqref{eq:parflow} in the parameter space with the vorticity
$\tilde{\omega}(\alpha,t)$.

As an example, we let $U(t)$ denote a three-dimensional spherical
shell region rotating with the Earth. A non-dimensional version of
the classical spherical parametrization of the globe is given by $\alpha=(\lambda/360^{\circ},\psi/360^{\circ},\rho/R)$,
where $\lambda$ and $\psi$ denote longitudes and latitudes in degrees,
$\rho$ denotes altitude in kilometers, in and $R$ denotes the radius
of the Earth in kilometers.

With the Earth modeled as a sphere placed of the $x$ coordinate system
and rotating with uniform angular velocity $\nu_{0}$ about the $x_{3}$
axis, we have the parametrization \eqref{eq:pardef} in the form 
\begin{equation}
f_{t}(\alpha)=R\left(\begin{array}{c}
\alpha_{3}\cos2\pi\alpha_{2}\cos\left(2\pi\alpha_{1}+\nu_{0}t\right)\\
\alpha_{3}\cos2\pi\alpha_{2}\sin\left(2\pi\alpha_{1}+\nu_{0}t\right)\\
\alpha_{3}\sin2\pi\alpha_{2}
\end{array}\right).\label{eq:parforearth}
\end{equation}

The velocity field $v(\alpha,t)$ in $U(t)$ and the corresponding
velocity field \eqref{eq:parflow} in the space $\mathcal{U}$ of
curvilinear coordinates are of the form 
\begin{equation}
v(\alpha,t)=\left(\begin{array}{c}
v_{\lambda}(\alpha,t)\\
v_{\psi}(\alpha,t)\\
v_{\rho}(\alpha,t)
\end{array}\right),\qquad\tilde{v}(\alpha,t)=\left(\begin{array}{c}
v_{\lambda}(\alpha,t)/\left[R\alpha_{3}\cos\left(2\pi\alpha_{1}+\nu_{0}t\right)\right]\\
v_{\psi}(\alpha,t)/\left(R\alpha_{3}\right)\\
v_{\rho}(\alpha,t)
\end{array}\right),\label{eq:upar}
\end{equation}
respectively, Here $(v_{\lambda},v_{\psi},v_{\rho})$ denote projections
of the velocity $v$ onto local unit vectors tangent to coordinate
lines of the latitude, longitude and altitude.

The parameter-space vorticity \eqref{eq:parvort} can be computed
from the velocity field $\tilde{v}(\alpha,t)$ in \eqref{eq:upar},
then used in computing the IVD and the LAVD. For the special case
of two-dimensional flows over a sphere, the $\rho$-component of $\tilde{v}$
is zero, and $\rho\equiv R$ in all formulas above.

\section{Appendix C: Proof of Theorem 2}

The linearization of \eqref{eq:parteq} along a particle motion $x_{p}(t)$
starting from a position $x_{0}$ at time $t_{0}$ gives the equation
of variations 
\[
\dot{\xi}=\left[\nabla v(x_{p}(t),t)+\tau(\delta-1)fJ\nabla v(x_{part}(t),t)+\mathcal{O}(\tau^{2})\right]\xi.
\]
By Liouville's theorem (Arnold 1978), the fundamental matrix solution
$P_{t_{0}}^{t}(x_{0})$ of this linear system of ODEs satisfies the
relationship 
\begin{eqnarray}
\det P_{t_{0}}^{t}(x_{0}) & = & e^{\int_{t_{0}}^{t}\mathrm{Trace}\left[\nabla v(x_{part}(s),s)+\tau(\delta-1)fJ\nabla v(x_{part}(s),s)+\mathcal{O}(\tau^{2})\right]ds}\nonumber \\
 & = & e^{\tau(\delta-1)f\int_{t_{0}}^{t}\omega_{3}(x_{part}(s),s)ds+\mathcal{O}(\tau^{2})}.\label{eq:detPform}
\end{eqnarray}

By smooth dependence of the solutions of \eqref{eq:parteq} on parameters,
over a finite time interval and for small enough $\tau$, the inertial
particle trajectory $x_{part}(s)$ is $\mathcal{O}(\tau)$ $C^{1}$-close
to the fluid particle trajectory $x(t;x_{0})$ starting from the same
initial position $x_{0}$ at time $t_{0}$. We thus have 
\[
\omega_{3}(x_{part}(s),s)=\omega_{3}(x(s;x_{0}),s)+\mathcal{O}(\tau),\qquad s\in[t_{0},t_{1}].
\]
Substituting this relation together with assumption \eqref{eq:cond1}
into \eqref{eq:detPform}, we obtain 
\begin{equation}
\det P_{t_{0}}^{t}(x_{0}^{*})=e^{\tau(\delta-1)f\int_{t_{0}}^{t}\omega_{3}(x(s;x_{0}^{*}),s)ds+\mathcal{O}(\tau^{2})}=e^{\tau(\delta-1)f\bar{\omega}_{3}(t)}e^{\mu\tau(\delta-1)f\mathrm{LAVD}_{t_{0}}^{t}(x_{0}^{*})+\mathcal{O}(\tau^{2})},\label{eq:det}
\end{equation}
with the two-dimensional LAVD field defined in \eqref{eq:LAVD2D},
and the sign parameter$\mu$ defined in \eqref{eq:cond1}. Note that
the planar form \eqref{eq:LAVD2D} of the LAVD applies in the present
spherical flow setting because the $\beta$-plane approximation is
assumed in the derivation of the reduced Maxey-Riley equation \eqref{eq:parteq}.

Over a finite time interval $[t_{0},t_{1}]$, no unique classical
attractor can be defined in a dynamical system. Indeed, by smooth
dependence on initial conditions, any attracting trajectory has an
open neighborhood filled with other attracting trajectories. What
prevails from such an open set as a uniquely observed finite-time
attractor is the trajectory that attracts the others at the strongest
rate.

Such a strongest attracting or repelling inertial particle motion
starting from an initial position $x_{0}$ is signaled by a local
extremum of the function $\det P_{t_{0}}^{t_{1}}(x_{0})$ at $x_{0}^{*}$,
i.e., by the relation $\nabla_{x_{0}}\left[\det P_{t_{0}}^{t_{1}}(x_{0}^{*})\right]=0$.
By \eqref{eq:det}, this extremum condition is equivalent to 
\[
e^{\mu\tau(\delta-1)f\mathrm{LAVD}_{t_{0}}^{t_{1}}(x_{0}^{*})}\left[\mu\tau(\delta-1)f\nabla_{x_{0}}\mathrm{LAVD}_{t_{0}}^{t_{1}}(x_{0}^{*})+\mathcal{O}(\tau^{2})\right]=0,
\]
which is in turn equivalent, for nonzero $\tau$, to an equation of
the general form 
\begin{equation}
\nabla_{x_{0}}\mathrm{LAVD}_{t_{0}}^{t_{1}}(x_{0}^{*})+\mathcal{O}(\tau)=0.\label{eq:imp}
\end{equation}

Assume now that $x_{0}^{*}$ is a non-degenerate maximum point of
$\mathrm{LAVD}_{t_{0}}^{t_{1}}(x_{0})$, i.e., we have 
\begin{equation}
\nabla_{x_{0}}\mathrm{LAVD}_{t_{0}}^{t_{1}}(x_{0}^{*})=0,\qquad\det\left[\nabla_{x_{0}}^{2}\mathrm{LAVD}_{t_{0}}^{t_{1}}(x_{0}^{*})\right]>0.\label{eq:corecond}
\end{equation}
Then, by the implicit function theorem, for small enough $\tau>0$,
the equation \eqref{eq:imp} has a unique solution of the form 
\begin{equation}
\bar{x}_{0}(\tau)=x_{0}^{*}+\mathcal{O}(\tau).\label{eq:close}
\end{equation}
Inertial particle trajectories starting from the initial position
$\bar{x}_{0}(\tau)$, therefore, prevail as the strongest finite-time
attractors or repellers over the time interval $[t_{0},t_{1}]$. These
attracting and repelling trajectories remain $\mathcal{O}(\tau)$
$C^{1}$- close to fluid particle trajectories starting from the positions
$x_{0}^{*}$.

Now, on a large enough domain, the spatially averaged relative vorticity
is approximately zero. (This already holds on the computational domain
used in Section \ref{sub:Two-dimensional-geostrophic-flow}). Thus,
the exponent in expression \eqref{eq:det} can be written as 
\[
\tau\left[\mu(\delta-1)f\mathrm{LAVD}_{t_{0}}^{t_{1}}(x_{0}^{*})+\mathcal{O}(\tau,f\bar{\omega}_{3})\right].
\]
When this expression is negative (positive) at the local maximum $x_{0}^{*}$
of $\mathrm{LAVD}_{t_{0}}^{t}$ , then the fluid trajectory starting
from $x_{0}^{*}$ approximates the locally strongest finite-time attractor
(repeller) of the inertial particle motion by \eqref{eq:close}. In
other words, when $\mu(\delta-1)f$ is negative (positive) at $x_{0}^{*}$,
the Lagrangian fluid trajectory starting from $x_{0}^{*}$ approximates
the locally strongest attractor (repeller) over the time interval
$[t_{0},t_{1}]$.

We conclude that in the limit of $\tau\to0$, cyclonic ($\mu f>0$)
Lagrangian eddy centers, as defined in Definition 1, are attractors
for light $(\delta>1$) particles and repellers for heavy particles
$(\delta<1$). Likewise, for $\tau\to0$, anticyclonic ($\mu f<0$)
Lagrangian eddy centers, as defined in Definition 1, are attractors
for heavy $(\delta<1$) particles and repellers for light $(\delta>1$)
particles. This completes the proof of Theorem 2.

\section{Appendix D: Comparison of LAVD-based vortex identification with other
objective approaches}

\subsection{Geodesic vortex detection}

Geodesic vortex detection seeks time $t_{0}$ positions of Lagrangian
vortex boundaries as outermost, closed stationary curves of the material-line-averaged
tangential stretching functional 
\[
Q(\gamma)=\frac{1}{\sigma}\int_{0}^{\sigma}\frac{\sqrt{\langle x_{0}^{\prime}(s),C_{t_{0}}^{t}(x_{0}(s))x_{0}^{\prime}(s)\rangle}}{\sqrt{\langle x_{0}^{\prime}(s),x_{0}^{\prime}(s)\rangle}}\,\mathrm{d}s,
\]
with $C_{t_{0}}^{t}(x_{0})=\left[F_{t_{0}}^{t}(x_{0})\right]^{T}F_{t_{0}}^{t}(x_{0})$
denoting the left Cauchy--Green strain tensor, and with $r(s)$ referring
to a parametrization of the closed curve $\gamma$ (cf. Haller \&
Beron--Vera 2013 and Haller 2015 for details). On such stationary
curves, the functional $Q$ must necessarily have a vanishing variation:
\begin{equation}
\delta Q(\gamma)=0.\label{eq:vari}
\end{equation}
This variational problem can be solved explicitly, with the solution
depending on the eigenvalues $\lambda_{i}(x_{0})$ and eigenvectors
$\xi_{i}(x_{0})$ of $C_{t_{0}}^{t}(x_{0})$, defined and indexed
as 
\[
C_{t_{0}}^{t}\xi_{i}=\lambda_{i}\xi_{i},\quad\left|\xi_{i}\right|=1,\quad i=1,2;\qquad0<\lambda_{1}\leq\lambda_{2},\qquad\xi_{1}\perp\xi_{2}.
\]

Using these quantities, all closed curves solving \eqref{eq:vari}
can be expressed as limit cycles of the autonomous differential equation
family 
\begin{equation}
x_{0}^{\prime}=\eta_{\lambda}^{\pm}(x_{0}),\qquad\eta_{\lambda}^{\pm}=\sqrt{\frac{\lambda_{2}-\lambda^{2}}{\lambda_{2}-\lambda_{1}}}\,\xi_{1}\pm\sqrt{\frac{\lambda^{2}-\lambda_{1}}{\lambda_{2}-\lambda_{1}}}\,\xi_{2},\label{eq:etafield}
\end{equation}
for some value of $\lambda>0$ and for some choice of the sign in
$\pm.$ This constant $\lambda$ turns out to be precisely the factor
by which any subset of a trajectory of \eqref{eq:etafield} will be
stretched under the flow map $\mathcal{\mathcal{F}}_{t_{0}}^{t}$.
Outermost members of nested limit cycle families of\eqref{eq:etafield}
are, therefore, locally the maximal closed curves in the flow that
stretch uniformly (i.e., without filamentation). The geodesic theory
of elliptic LCSs developed by Haller \& Beron--Vera (2013) defines
coherent Lagrangian vortex boundaries to be these outermost limit
cycles. These boundaries are objective by the objectivity of the invariants
of $C_{t_{0}}^{t}(x_{0})$. An automated detection algorithm for geodesic
vortex boundaries is given by Karrasch et al. (2014). Geodesic vortex
detection has no direct extension to three-dimensional flows, but
a variational approach for nearly-uniformly-stretching material surfaces
is available (\"{O}ttinger et al. 2015).

\subsection{Ellipticity--time diagnostic}

Haller (2001) studies the finite-time stability of a fluid trajectory
$x(t;t_{0},x_{0})$ in a frame aligned with to the eigenvectors of
the rate-of-strain tensor $D(x,t)$ along the trajectory. A topological
argument shows that the trajectory has an \emph{instantaneous elliptic
stability} type if throughout the time interval of interest, either
$D$ vanishes or the strain-acceleration tensor 
\begin{equation}
M=\dot{D}+2D\nabla v,\label{eq:strain_acceleration}
\end{equation}
is indefinite on the zero rate of strain set $Z=\left\{ a\in\mathbb{R}^{2}:\,\left\langle a,Da\right\rangle =0\right\} $.
(In \eqref{eq:strain_acceleration}, dot refers to the material derivative.)
The ellipticity time $\tau_{e}(t_{1},t_{0},x_{0})$ for a trajectory
released from $x_{0}$ at time $t_{0}$ is then defined as the percentage
of time within $[t_{0},t_{1}]$ over which the trajectory has instantaneous
elliptic stability.

The scalar field $\tau_{e}(t_{1},t_{0},x_{0})$ is an objective, pointwise
indicator of fluid trajectory stability. It does not offer a strict
definition of a material vortex boundary, but its high values indicate
the general location of material vortices. As shown in Haller (2001),
$\tau_{e}(t_{1},t_{0},x_{0})$ can equivalently be defined as the
percentage of time over which the trajectory $x(t;t_{0},x_{0})$ is
elliptic in the sense of the Okubo--Weiss criterion, applied in a
frame co-rotating rate-of-strain eigenbasis. A similar ellipticity
time diagnostic can be defined in three dimensions (Haller 2005),
but this extension is no longer related to other known instantaneous
vortex criteria.

To be experimentally verifiable, a vortex criterion based on coherent
rotation of fluid elements should have a direct relation to observable
mean material rotation visualized by small inertial tracers with an
attached arrow (vorticity meters). Experiments show that this mean
material rotation has an angular velocity that is precisely one half
of the local vorticity (Shapiro 1967). In contrast, the relative vorticity
observed in strain basis theoretical remains invisible under all possible
Euclidean observer changes from the lab frame. Indeed, the coordinate
change to the pointwise differing rate-of-strain eigenbases would
require a spatially nonlinear rotation tensor $Q(x,t)$ in \eqref{eq:observer_change},
under which \eqref{eq:observer_change} no longer describes a physically
meaningful observer change.

\subsection{Comparison with LAVD on the Agulhas leakage data set}

Both the geodesic and the ellipticity-time approach require more computational
effort than the LAVD approach. For the geodesic method, limit-cycle
families of a vector field composed of the invariants of the Cauchy--Green
strain tensor must be computed with high accuracy, requiring the accurate
differentiation of trajectories with respect to their initial conditions.
For the ellipticity-time approach, the time-derivative of the rate-of-strain
eigenbasis must be determined along trajectories wit high accuracy.

Both approaches are also more stringent than the LAVD approach, requiring
uniform stretching (geodesic method) or a lack of trajectory-level
instability (ellipticity-time diagnostic) along the Lagrangian vortex
boundaries. By their objectivity, both approaches should generally
capture the same Lagrangian vortex region as the LAVD approach, but
are expected to yield tighter vortex boundaries because of their more
stringent definitions of coherence and trajectory stability, respectively.

\begin{figure}
\centering
\subfloat[]{\includegraphics[width=0.5\textwidth]{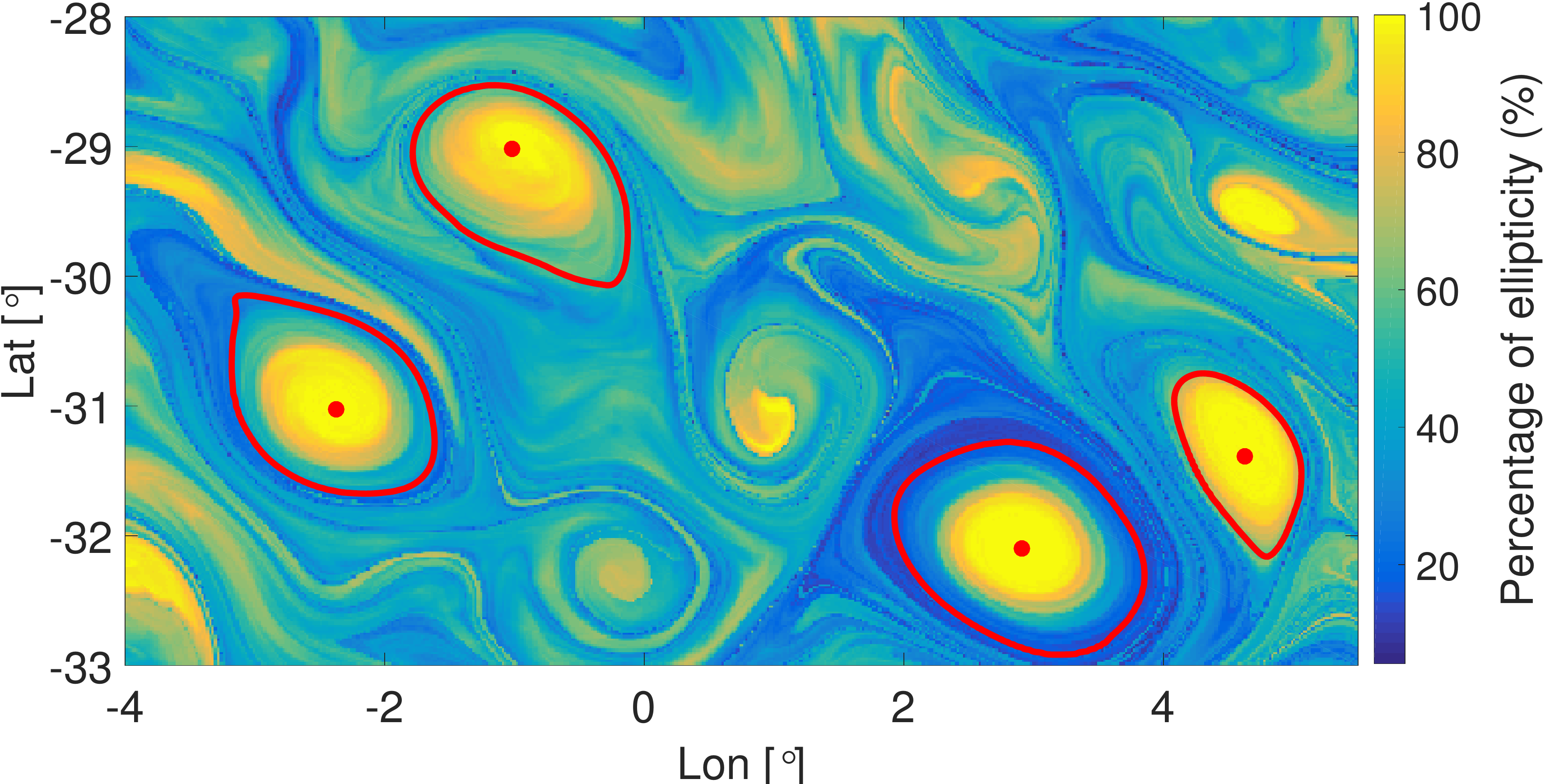}}\quad
\subfloat[]{\includegraphics[width=0.43\textwidth]{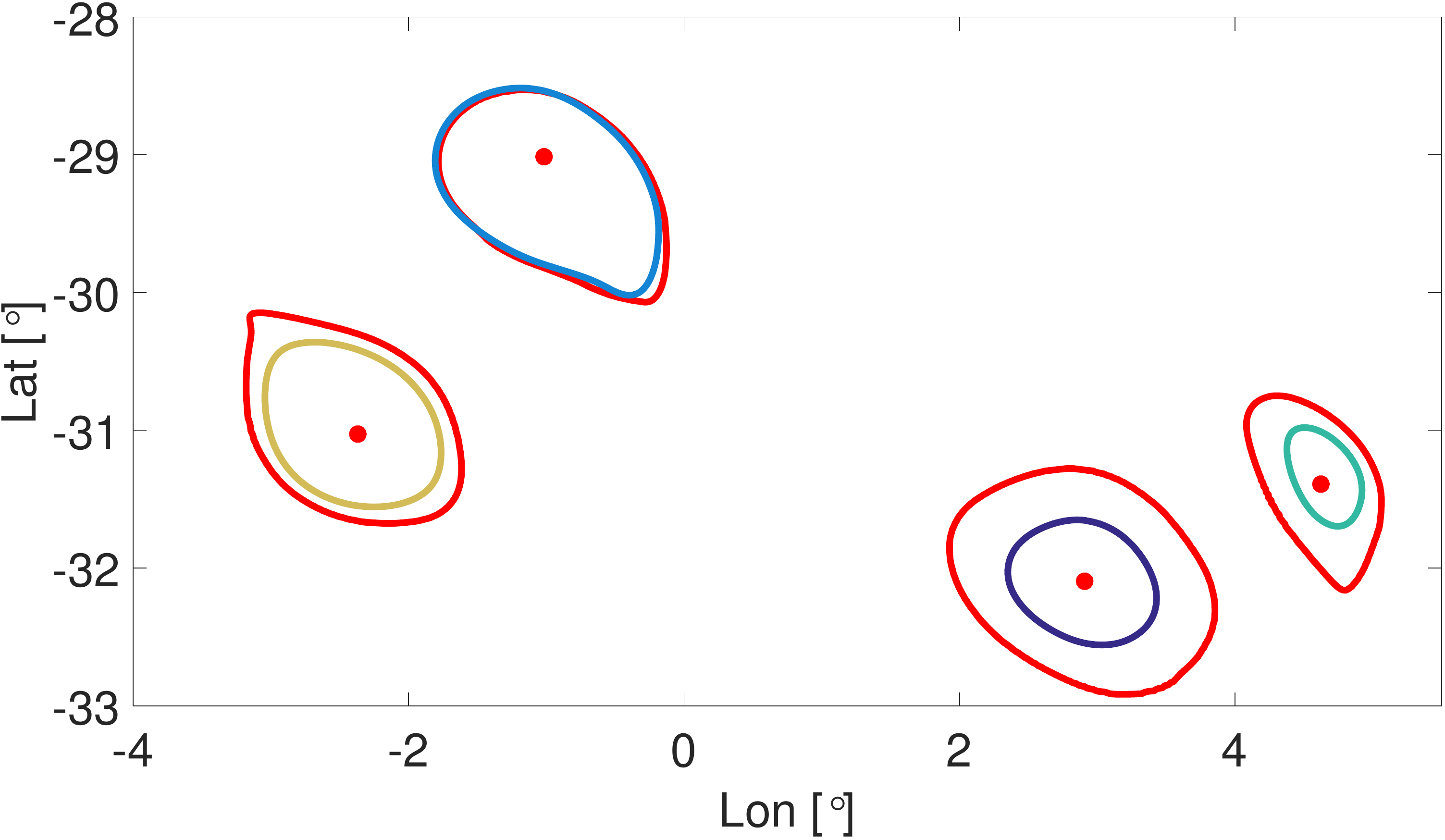}}
\caption{(a) Ellipticity time field $\tau_{e}(t_{1},t_{0},x_{0})$ for the
Agulhas leakage data set of Section 10.6 with $t_{0}=\mathrm{November}\,11,\,2006$
and $t_{1}=t_{0}+90\,\mathrm{days}$. Rotationally coherent Lagrangian
vortex boundaries (red) are superimposed for reference. (b) Geodesic
vortex boundaries at time $t_{0}$ for the same data set, with the
LAVD-based vortex boundaries (red) superimposed.}
\label{fig:comparison} 
\end{figure}

Figure \ref{fig:comparison} shows the results from these two alternative
approaches for the data set analyzed in section 10.5, with the LAVD-based
vortex boundaries superimposed in red. For the purposes of this comparison, we have used the numerical implementation of the geodesic eddy detection method described by Hadjighasem \& Haller (2016).

As seen in Fig. \ref{fig:comparison}a, the pointwise ellipticity
time diagnostic highlights the same material vortex regions labelled
as coherent by the other two methods. However, it also suggests further
vortical regions that are neither rotationally no stretching-wise
coherent. The diagnostic does not offer a well-defined boundary for
the detected vortices either.

Fig. \ref{fig:comparison}b confirms the expectation that the variationally
derived geodesic vortex detection method generally labels the same
coherent material vortices, but yields smaller vortex boundaries due
to its uniform stretching requirement. This, coupled with the significantly
decreased computational cost and coding effort, renders LAVD-based
Lagrangian vortex identification preferable over the other two objective
methods considered here.

\newpage{}

\end{document}